\documentclass[11pt]{article}

\usepackage{booktabs}
\usepackage{multirow}
\usepackage{enumitem}
\usepackage{adjustbox}
\usepackage[table]{xcolor}
\usepackage{array}
\usepackage[normalem]{ulem}
\usepackage{siunitx}
\usepackage{etoolbox}

\usepackage{tikz}

\definecolor{OBBlue}{HTML}{F0F6FF}
\definecolor{AvgGray}{HTML}{F8F9FA}

\newcommand{\sig}[1]{\ifx\relax#1\relax\else\rlap{\textsuperscript{\tiny #1}}\fi}
\newcommand{\best}[2]{\bfseries #1\sig{#2}}
\newcommand{\sbest}[2]{\itshape #1\sig{#2}}
\DeclareRobustCommand{\GACL}{GACL}
\newcommand{\Nout}{N_{\mathrm{out}}}
\newcommand{\RA}{R_A}
\newcommand{\MA}{M_A}
\newcommand{\MJ}{M_J}
\newcommand{\alphamax}{\alpha_{\max}}

\usepackage[preprint]{acl}

\DeclareRobustCommand{\figref}[1]{\hyperref[#1]{Figure~\ref*{#1}}}
\DeclareRobustCommand{\tabref}[1]{\hyperref[#1]{Table~\ref*{#1}}}
\DeclareRobustCommand{\appref}[1]{\hyperref[#1]{Appendix~\ref*{#1}}}
\DeclareRobustCommand{\secref}[1]{\hyperref[#1]{\S\ref*{#1}}}
\providecommand{\linenumbers}{}
\providecommand{\nolinenumbers}{}
\makeatletter
\@ifundefined{linenumbersep}{\newlength{\linenumbersep}}{}
\makeatother
\newif\ifbodylineno
\bodylinenotrue
\pretocmd{\section}{\ifbodylineno\linenumbers\fi}{}{}
\newcommand{\suspendbodylineno}{\ifbodylineno\nolinenumbers\fi}
\newcommand{\resumebodylineno}{\ifbodylineno\linenumbers\fi}
\AtBeginEnvironment{figure}{\suspendbodylineno}
\AfterEndEnvironment{figure}{\resumebodylineno}
\AtBeginEnvironment{figure*}{\suspendbodylineno}
\AfterEndEnvironment{figure*}{\resumebodylineno}
\AtBeginEnvironment{table}{\suspendbodylineno}
\AfterEndEnvironment{table}{\resumebodylineno}
\AtBeginEnvironment{table*}{\suspendbodylineno}
\AfterEndEnvironment{table*}{\resumebodylineno}
\usepackage[T1]{fontenc}
\usepackage{newtxtext}
\usepackage{latexsym}
\usepackage{microtype}
\usepackage{graphicx}
\usepackage{amsmath}
\usepackage{amssymb}
\usepackage{tabularx}
\usepackage{float}
\usepackage{caption}
\usepackage{longtable}
\usepackage[most]{tcolorbox}
\usetikzlibrary{arrows.meta,positioning,calc}
\renewcommand{\abstractname}{Abstract}
\newtcolorbox{promptbox}[1][]{enhanced,breakable,colback=gray!2,colframe=black!65,colbacktitle=black!75,coltitle=white,boxrule=0.5pt,arc=2.2mm,outer arc=2.2mm,left=1.6mm,right=1.6mm,top=1.1mm,bottom=1.1mm,toptitle=0.8mm,bottomtitle=0.8mm,fonttitle=\bfseries\small,#1}
\newtcolorbox{casebox}[1][]{enhanced,breakable,colback=gray!2,colframe=black!35,boxrule=0.4pt,arc=1.5mm,left=1.5mm,right=1.5mm,top=1mm,bottom=1mm,fonttitle=\bfseries,#1}
\renewcommand{\arraystretch}{1.15}
\title{Beyond Text Following: Repairable Arbitration Reversals in Audio-Language Models}
\author{
\textbf{Yichen Gao$^{1}$, Yiqun Zhang$^{1,2}$, Zijing Wang$^{1}$, Yujia Li$^{1}$, Heng Guo$^{1}$,}\\
\textbf{Xi Wu$^{1}$, Xiaocui Yang$^{1}$, Shi Feng$^{1}$, Yifei Zhang$^{1}$, Daling Wang$^{1}$}\\
$^1$Northeastern University, China; $^2$Shanghai Artificial Intelligence Laboratory, China\\
\texttt{\{gaoinfinity1,zhangyiqun344,wzj1718\}@gmail.com}\\
\texttt{\{liyujia,guoh4,wux2\}@mails.neu.edu.cn}\\
\texttt{\{yangxiaocui,fengshi,zhangyifei1,wangdaling\}@cse.neu.edu.cn}
}
\date{}
\begin{document}
\raggedbottom
\hbadness=10000
\hfuzz=60pt
\maketitle
\setcounter{figure}{0}
\begin{abstract}
Audio-language models (ALMs) often follow text that conflicts with audio, even when the audio evidence is clear. This raises a basic question: is the audio-supported answer unavailable, or is it represented but overridden by the conflicting text? We examine this question using a same-audio counterfactual that keeps the audio fixed, removes only the conflicting text, and measures the resulting shift in model preference. Across five ALMs and four conflict tasks, 64.1\% of conflict samples show a sign flip: the same-audio branch prefers the audio-supported answer, whereas the joint branch prefers the text-supported answer. This pattern suggests that the relevant audio evidence is encoded but loses in arbitration. Activation patching further localizes the reversal to answer-position computation, and patching effects closely track output candidate-score differences (Spearman $\rho{=}0.93$). Using this diagnostic, we propose \textbf{G}ated \textbf{A}udio \textbf{C}ounterfactual \textbf{L}ogit Correction (\GACL{}), a training-free decoding rule that interpolates between joint and same-audio scores. Under a strict 5,pp faithfulness-drop budget, \GACL{} improves nAUC by 17.8 points over the best contrastive baseline and transfers without retuning to vision--text arbitration (up to +40.5,pp).
\end{abstract}
\section{Introduction}

\begin{figure}[t]
\centering
\includegraphics[width=\columnwidth]{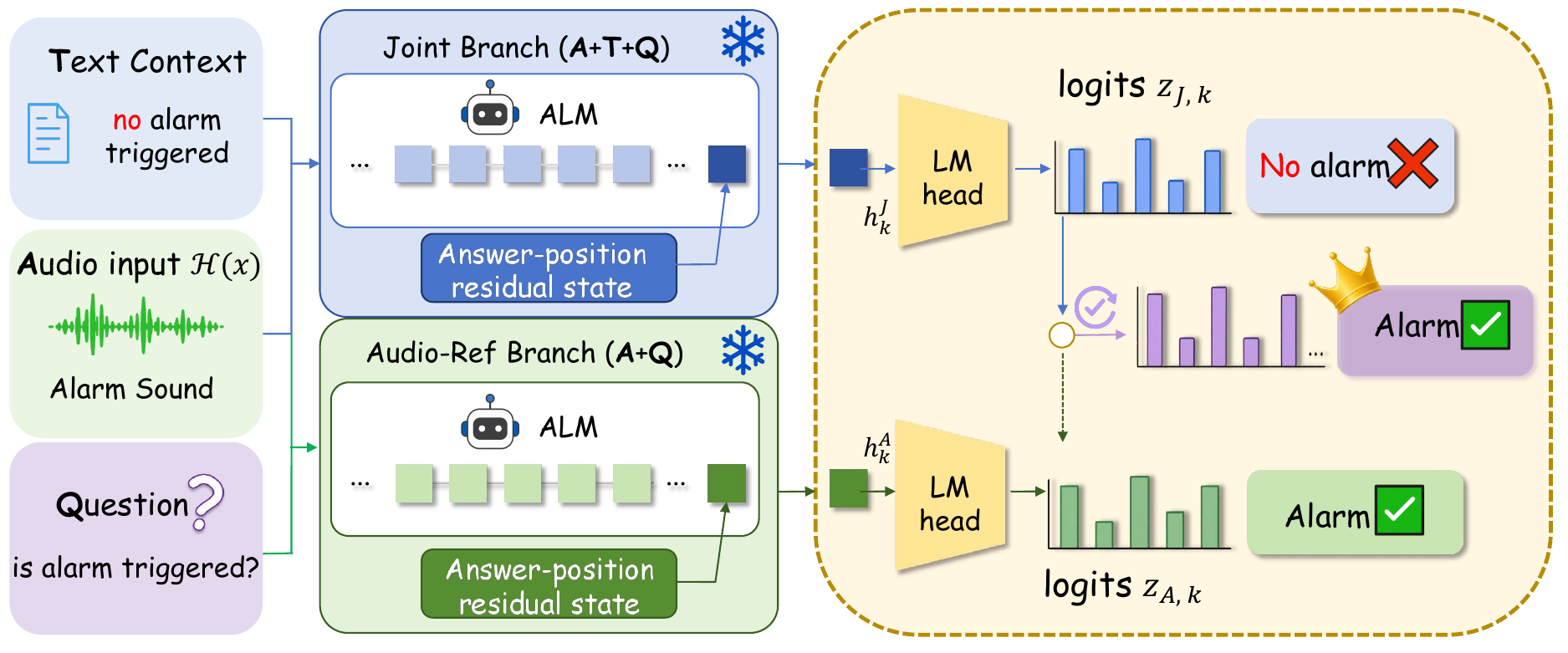}
\caption{\textbf{GACL overview.}
The joint branch conditions on audio plus conflicting text and predicts the text-supported answer; the audio-reference branch removes only the text and predicts the audio-supported answer. GACL freezes both ALM branches and applies gated, bounded interpolation from joint logits toward reference logits.}
\label{fig:schema}
\end{figure}

When an audio-language model (ALM) hears an alarm while the text states that no alarm was triggered, it may answer ``no alarm.'' This error shows text following, but not how the decision was made. The model may have missed the alarm, or encoded it and let the textual claim override it. This
distinction matters: perceptual failure calls for better acoustic
encoding, whereas arbitration failure calls for better conflict
resolution. As ALMs~\citep{qwen2audio,qwen25omni,liu2025voxtral} move beyond
transcription into agentic settings such as meeting assistance and
emergency triage, they may receive audio together with written
context such as agendas, logs, or incident notes. When that context
conflicts with the audio, users need models that resolve the disagreement
based on the heard evidence rather than repeating the text.

Recent benchmarks expose this symptom but not its mechanism.
MCR-Bench~\citep{mcrbench} and ALME~\citep{alme} show that ALMs
often choose the text-supported answer when audio and text conflict,
when the audio evidence is clear. These benchmarks, however, do not show
whether the audio-supported answer is unavailable or whether it remains
available but loses to the text during arbitration. Decoding-time
methods face the same limitation: their reference branches weaken
modality evidence. Audio methods such as AAD and
ACD~\citep{aad,acd}, together with their vision--language counterparts
VCD, OPERA, and ICD~\citep{vcd,opera,icd}, construct reference branches by
weakening or removing modality evidence. Such references can diagnose
reliance on degraded evidence, but audio--text arbitration requires a
different counterfactual: with the same audio held fixed, how does model
preference change when only the conflicting text is removed?

We study this counterfactual directly. For each sample, we keep the audio unchanged, remove the conflicting text, and
compare preferences for the audio-supported and text-supported
answers. Across five ALMs and four conflict tasks, 64.1\% of conflict
samples show a sign flip: the same-audio branch prefers the
audio-supported answer, whereas the joint audio--text branch prefers the
text-supported answer. The audio answer therefore remains available, but
it loses once the text is added. We refer to this pattern as
\emph{repairable\footnote{\emph{Repairable} is used operationally: samples
in $O$ have an audio-favoring same-audio reference and a text-favoring
joint branch, providing a candidate set for correction.} arbitration
reversal}. Activation patching further localizes the reversal to
the answer-position residual stream, and the patch-induced direction
tracks the score difference $s_A - s_J$ at the rank level
($\rho{=}0.93$). This alignment makes the repair direction visible from
outputs alone, and motivates a decoding-time fix.

Using this diagnostic, we propose \textbf{G}ated \textbf{A}udio \textbf{C}ounterfactual \textbf{L}ogit
Correction (\GACL{}; \figref{fig:schema}), a training-free decoding rule that interpolates
between joint and same-audio reference scores, gated by branch
disagreement and reference reliability. Under a 5\,pp faithfulness-drop
budget, \GACL{} improves conflict audio-following accuracy by
17.8 nAUC points over the best contrastive decoding baseline. Applied
without retuning to vision--text arbitration on MC$^2$~\citep{mc2}, the
same rule raises adversarial accuracy by up to 40.5\,pp, suggesting that
this counterfactual design generalizes beyond audio--text.

Our key contributions are:
\begin{itemize}[nolistsep, noitemsep, leftmargin=*]
  \item We identify repairable arbitration reversal as a common
        recoverable failure mode in audio--text conflict: models can
        support the audio answer, but the joint input reverses the
        decision.
  \item We localize this reversal to the answer-position residual stream
        and show that the patch-induced direction tracks the output
        score difference $s_A - s_J$ at the rank level (Spearman
        $\rho{=}0.93$), making the repair direction visible from outputs
        alone.
  \item We propose \GACL{}, a diagnostic-derived decoding rule improving audio--text rescue by 17.8 nAUC under a 5\,pp faithfulness budget and generalizing to vision--text conflict without retuning.
\end{itemize}
\section{Related Work}
\paragraph{Audio-language models and evidence conflict.}
Recent ALMs adapt instruction-following LLMs to speech, music, and acoustic events via audio encoders or unified multimodal tokenization~\citep{salmonn,ltu,gama,audio_flamingo,qwen2audio,qwen25omni,qwen3omni,kimi_audio,liu2025voxtral}. Beyond transcription, text can become competing evidence rather than merely a task cue. MCR-Bench~\citep{mcrbench} and ALME~\citep{alme} show that mainstream ALMs often follow text under explicit conflict. This behavior is related to evidence-arbitration failures studied in knowledge conflicts, context faithfulness, and sycophancy-like instruction following~\citep{knowledge_conflicts,context_faithfulness,sharma_sycophancy,wei_sycophancy}. These benchmarks identify the failure mode, but not whether the audio-supported answer is unavailable or instead remains available but is displaced at the decision point.

\paragraph{Reference design in decoding-time correction.}
Decoding-time correction methods adjust predictions using reference distributions, perturbed inputs, or penalties, without parameter updates. The reference defines the counterfactual error the method can suppress. Model-level contrast~\citep{contrastive_decoding}, layer-level contrast~\citep{dola}, vision--language methods~\citep{vcd,icd,opera}, and audio contrastive methods such as AAD and ACD~\citep{aad,acd} typically weaken, remove, or perturb modality evidence. These references measure reliance on degraded evidence, but arbitration requires a different test: with the same audio held fixed, which answer does the model support after only the competing text is removed? This gap in reference design motivates our direct comparison with degraded-reference baselines under a rescue--faithfulness trade-off.
\section{Diagnosis: From Behavior to a Repairable Signature}
\label{sec:setup}
When an ALM follows text in an audio--text conflict, two failures can look identical from output alone. The model may fail to extract the relevant audio evidence, or extract it but still allow the text to determine the answer. These cases require different repairs. We therefore begin with a counterfactual comparison that changes only one factor: whether the conflicting text is present.
\subsection{Setup: Two Branches and Two Margins}
\label{sec:diagnosis-setup}
\paragraph{Two branches.}
For each conflict instance, we use two prompts with the same audio $x$, question $q$, candidate set $\mathcal{C}$, and output format. They differ only in whether the conflicting textual description is included (\figref{fig:two-branch}):
\begin{figure}[t]
\centering
\includegraphics[width=0.8\columnwidth]{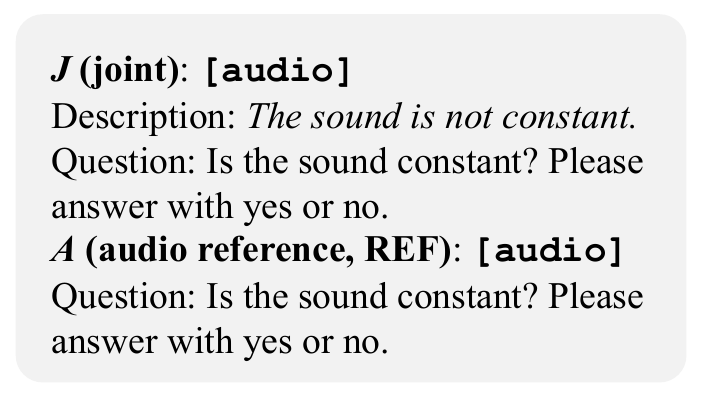}
\caption{\textbf{Two-branch setup.} The joint branch ($J$) conditions on audio plus conflicting text; the same-audio reference ($A$) removes only the text. Both share the same audio, question, and output format.}
\label{fig:two-branch}
\end{figure}
Here, the audio contains a continuous 23\,s alarm, so the audio-supported answer is $y_a=\text{yes}$, while the conflicting text supports $y_t=\text{no}$. Because $J$ and $A$ keep the audio, question, candidate set, and output format fixed, the comparison isolates the effect of adding the conflicting text.
\paragraph{Two margins.}
A top-1 prediction hides preference strength and direction. We therefore use signed log-probability margins. For each branch $b\in\{A,J\}$, we score closed-set verbalizers and write $s_b(c)$ for the length-normalized, candidate-normalized log-probability of candidate $c\in\mathcal{C}$. \appref{app:prompt-templates} and \appref{app:scoring-validation} give the scoring details. We define
\begin{equation}
\MA = s_A(y_a)-s_A(y_t),
\MJ = s_J(y_a)-s_J(y_t).
\label{eq:margins}
\end{equation}
These margins answer two distinct questions. $\MA>0$ indicates that, without the conflicting text, the model prefers the audio-supported answer, testing whether the audio branch can support that answer. $\MJ<0$ indicates that, after the conflicting text is added, the model prefers the text-supported answer; this tests whether arbitration fails. Because both margins are computed on the same candidate pair $(y_a,y_t)$, they can be compared directly.
The diagnostic region is
\begin{equation}
O=\{\MA>0,\,\MJ<0\}.
\label{eq:O}
\end{equation}
Samples in $O$ exhibit the behavioral signature of arbitration failure: the audio answer is available in the same-audio reference branch, but the joint branch flips toward the text-supported answer. We treat $O$ as a diagnostic region for possible arbitration failure, not a claim that every sample in this region can be repaired by a particular method.
\paragraph{Models and datasets.}
We evaluate five open-weight audio-language/omni-modal models (7B--30B parameters): Qwen2-Audio-7B-Instruct~\citep{qwen2audio}, Qwen2.5-Omni-7B~\citep{qwen25omni}, Voxtral-Small-24B~\citep{liu2025voxtral}, Qwen3-Omni-30B-A3B-Instruct~\citep{qwen3omni}, and Kimi-Audio-7B-Instruct~\citep{kimi_audio}. We use four audio--text conflict tasks: AQA (Audio Question Answering), VSC (Visual Scene Classification), and SER (Speech Emotion Recognition) from MCR-Bench~\citep{mcrbench} for audio--description conflict, and the English subset of ALME (Audio-LLM Modality Evaluation)~\citep{alme} for speech--transcript conflict. Each provides 200 development examples for hyperparameter selection plus a held-out 200-example test split. Appendix~\ref{app:models-benchmarks} gives details.
\subsection{Behavioral Evidence}
\label{sec:behavioral-evidence}
\begin{figure*}[!htbp]
\centering
\includegraphics[width=\textwidth]{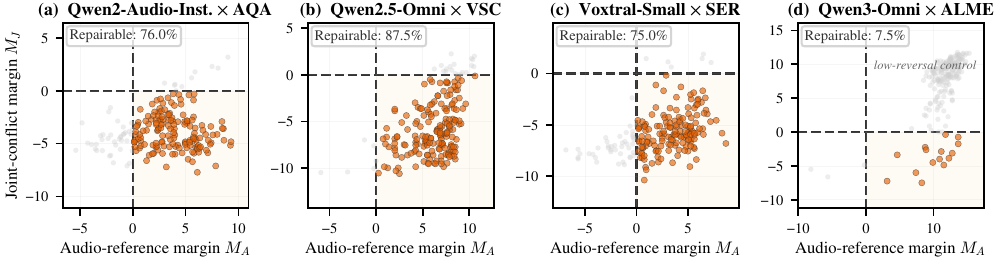}
\caption{\textbf{Conflict failures show up as a signed margin reversal.} $\MA$ is on the $x$-axis and $\MJ$ on the $y$-axis. Orange points mark samples in $O=\{\MA>0,\MJ<0\}$, where the same candidate pair is audio-favored in the same-audio reference branch but text-favored in the joint branch. \appref{app:margin-grid} shows the full $5\times4$ grid.}
\label{fig:2}
\end{figure*}

We now apply the two-branch diagnostic to the conflict benchmarks. The goal is to measure conflict failure and determine whether it reflects missing audio evidence or arbitration reversal. We compare three conditions: faithful inputs, where the accompanying text, if present, is consistent with the audio-supported answer; the joint conflict branch $J$, where the conflicting text is present; and the same-audio reference branch $A$, where the conflicting text is removed while the audio is kept fixed. \tabref{tab:1} summarizes top-1 behavior, and \figref{fig:2} shows the signed margins on conflict samples.
Faithful inputs reach 97.9\% audio-following; conflict drops accuracy to 18.0\% with 80.3\% of predictions matching the text---a large, systematic shift from the audio.
Yet when only the conflicting text is removed, the $A$-branch gives the audio-supported answer on 76.6\% of conflict samples---the audio channel still carries the information.
The margin plot further shows that the dominant pattern is signed reversal, not a small confidence decrease. If conflict merely weakened preference for $y_a$, many samples would lie near $\MJ=0$. Instead, 64.1\% of conflict samples fall in $O=\{\MA>0,\MJ<0\}$: the same candidate pair is audio-favored in $A$ but text-favored in $J$. This sign flip provides the behavioral basis for the internal analysis in the next section.
The $A$-branch is informative on AQA/VSC/ALME (75.9--96.5\%) but weak on SER (48.9\%); we treat this as a benchmark property and revisit its consequences in \secref{sec:component}.
\begin{table}[t]
\centering
\caption{\textbf{Behavioral evidence for arbitration failure.} Faithful and Conflict columns report audio-following accuracy; Text-follow reports the fraction of conflict predictions matching $y_t$; Audio-ref reports accuracy after removing only the conflicting text; $O$ is the share of conflict samples in $\{\MA>0,\MJ<0\}$. Results are macro-averaged across five models. Per-configuration results appear in \appref{app:per-config-behavioral}.}
\label{tab:1}
\footnotesize
\begin{adjustbox}{max width=\columnwidth}
\begin{tabular}{lccccc}
\toprule
Task & Faithful & Conflict & Text-follow & Audio-ref & $O$ \\
\midrule
AQA & 99.4 & 18.2 & 81.8 & 85.0 & 69.3 \\
VSC & 96.7 & 14.4 & 81.6 & 75.9 & 73.1 \\
\rowcolor{AvgGray} SER & 98.4 & 5.0 & 92.0 & 48.9 & 56.6 \\
ALME & 97.1 & 34.4 & 65.6 & 96.5 & 57.5 \\
\midrule
\rowcolor{AvgGray} Macro & 97.9 & 18.0 & 80.3 & 76.6 & 64.1 \\
\bottomrule
\end{tabular}
\end{adjustbox}
\end{table}
The behavioral evidence suggests that many conflict errors are not reducible to perceptual failure: removing only the conflicting text often recovers the audio-supported answer, and many samples show a signed reversal from audio-favored in $A$ to text-favored in $J$. However, behavior alone does not reveal the joint branch's internal computation. One possibility is that the joint branch never makes audio-compatible evidence available at the answer position, while the same-audio branch reaches $y_a$ through a different computation path. Another possibility is that the joint branch contains audio-compatible evidence, but this evidence is suppressed or overwritten during readout. These two accounts imply different repair targets, so we turn to the model's internal computation.

\section{From Mechanism to a Decoding Rule}
\label{sec:loc-output}

\secref{sec:setup} left one question open: in $J$, does the model compute $y_a$ and then let text dominate at readout, or fail to compute $y_a$ at all? The answer determines where repair should intervene. We ask two questions: can the difference between $J$ and $A$ be causally localized inside the model, and does it appear in the final candidate scores, the only quantities available to an output-space rule?

\subsection{Causal Localization via Activation Patching}
\label{sec:loc-residual}

We use activation patching, a standard causal test for whether an internal state drives an output~\citep{vig2020causal,meng2022rome,activation_patching}. We inject hidden states from $A$ into matched positions of $J$ and test whether $J$ returns to $y_a$. Since $A$ and $J$ differ in prompt length, we match positions by semantic role rather than absolute index: audio-span states map to the shared audio segment, instruction-token states to the shared question and output format, and answer-position states to the final prompt position before generation. We patch the residual stream as the primary site, since it aggregates component outputs.

Patching requires a clean source--target direction, so we restrict to the reference-correct subset $R=\{\hat y_A=y_a,\,\hat y_J\neq y_a,\,x\in O\}\subseteq O$. On $R$, we inject residual states from $A$ into $J$ at the three position classes and measure rescue rates in \tabref{tab:2}.
\emph{Answer-position residual} patching reaches a macro average of 0.81. Replacing $J$'s state at the answer commitment point recovers $y_a$ on most samples. By contrast, \emph{audio-span} and \emph{instruction-token} residual patching give zero rescue in the reported controls. Because the branches share audio and instructions, these sites serve as negative controls. Their zero rescue suggests that the relevant difference is not in perception or instruction handling, but downstream near response generation.

\begin{table}[t]
\centering
\caption{\textbf{The repair direction localizes to the answer boundary.}
Answer-side residual patching recovers the audio-supported answer across models and tasks, whereas audio-span and instruction-token interventions give zero rescue in all controls.}
\label{tab:2}
\footnotesize
\setlength{\tabcolsep}{4.0pt}
\renewcommand{\arraystretch}{1.08}
\begin{adjustbox}{max width=\columnwidth}
\begin{tabular}{@{}lccccc@{}}
\toprule
\textbf{Site / Model} & \textbf{AQA} & \textbf{VSC} & \textbf{SER} & \textbf{ALME} & \textbf{Avg.} \\
\midrule
\multicolumn{6}{@{}l}{\textit{Answer-side residual}} \\
Qwen2-Audio-Inst. & 1.00 & 0.96 & 0.44 & 0.81 & 0.80 \\
Qwen2.5-Omni      & 1.00 & 0.88 & 0.54 & 0.86 & 0.82 \\
Voxtral-Small     & 1.00 & 0.95 & 0.56 & 0.87 & 0.85 \\
Qwen3-Omni        & 0.94 & 0.88 & 0.58 & 0.98 & 0.85 \\
Kimi-Audio-Inst.  & 0.94 & 0.62 & 0.40 & 0.90 & 0.72 \\
\addlinespace[1pt]
\rowcolor{AvgGray}
\textbf{Macro} & \textbf{0.98} & \textbf{0.86} & \textbf{0.50} & \textbf{0.88} & \textbf{0.81} \\
\midrule
\multicolumn{6}{@{}l}{\textit{Negative controls}} \\
Audio-span residual        & 0.00 & 0.00 & 0.00 & 0.00 & 0.00 \\
Instruction-token residual & 0.00 & 0.00 & 0.00 & 0.00 & 0.00 \\
\bottomrule
\end{tabular}
\end{adjustbox}
\end{table}

Two additional checks refine this picture. Component-level patching shows that neither attention nor the MLP alone explains the full residual effect; the failure is better described as a distributed residual-stream write than as a single-module artifact (\appref{app:component-decomposition}). Layer-wise sweeps place the rescue effect in a mid-to-late layer band, where the model appears to move from evidence integration to answer commitment; we call this band the \emph{commit window} and define it operationally in \appref{app:bidirectional-sweeps}. A linear-probing audit~\citep{alain_bengio_probes,hewitt_liang_2019} on Qwen2-Audio-Inst.$\times$VSC gives the same interpretation: $y_a$ remains decodable from late states of both branches ($\sim$0.89), but cross-branch decoding collapses after the commit window (to 0.087). The audio answer is not erased; the readout direction changes (\appref{app:logit-lens-probing}).

In $J$, the audio answer is computed and remains accessible; conflict changes the answer-side readout. The failure is readout, not perception.

\subsection{Bridging Internal Repair Direction to Output Scores}
\label{sec:bridge}

A hidden-state repair direction does not imply that an output-space rule can use it. A patching effect is local to the injection site, whereas the final score difference between branches includes all downstream computation. These directions could therefore differ.

\begin{figure*}[t]
\centering
\includegraphics[width=0.80\textwidth]{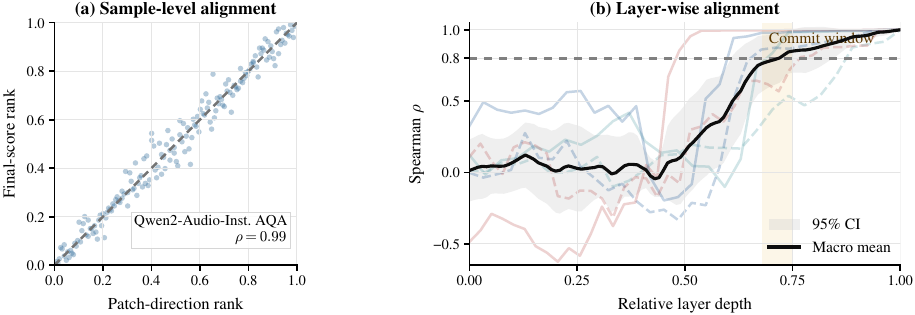}
\caption{\textbf{The internal repair direction is visible in final scores.} \textbf{(a)} Per-sample rank alignment between the patch-induced direction and $s_A-s_J$ on Qwen2-Audio-Inst.$\times$AQA (macro across twelve configurations $\rho{=}0.93$), supporting $s_A-s_J$ as a training-free proxy for the hidden repair direction. \textbf{(b)} Layer-resolved alignment starts near zero and rises through the commit window (shaded). \appref{app:mechanism-evidence} gives per-configuration curves.}
\label{fig:3}
\end{figure*}

We test alignment directly. For each sample, we compute the patch-induced displacement at the best commit-window layer and the observable candidate-level difference $s_A-s_J$. We compare them with Spearman $\rho$ (\figref{fig:3}). Across twelve model--task configurations, macro alignment is $\rho=0.93$, well above the raw logit-lens projection~\citep{logit_lens} ($\rho=0.67$). Using $s_A-s_J$ as the rescue direction reaches 92.3\% best rescue, compared with 54.7\% for the lens. Across layers, alignment is near zero early and rises through the same commit window where patching becomes effective (\figref{fig:3}b).
\begin{table*}[t]
\centering
\caption{\textbf{Main results.} nAUC (\%, higher is better) under strict 0--5\,pp and relaxed 0--10\,pp faithfulness budgets. \textbf{Bold} marks the best within each model and metric; \emph{italics} mark second-best. Superscripts denote paired-bootstrap significance of \GACL{} over the stronger of AAD/ACD on the same row (* $p<0.05$, ** $p<0.01$, *** $p<0.001$).}
\label{tab:main-results}
\footnotesize
\setlength{\tabcolsep}{3.0pt}
\begin{adjustbox}{max width=\textwidth}
\begin{tabular}{
@{}ll
*{8}{S[table-format=2.1,table-space-text-post={***}]}
@{\hspace{8pt}}
>{\columncolor{AvgGray}}S[table-format=2.1,table-space-text-post={***}]
>{\columncolor{AvgGray}}S[table-format=2.1,table-space-text-post={***}]
@{}}
\toprule
\multirow{2}{*}{Model}
& \multirow{2}{*}{Method}
& \multicolumn{2}{c}{AQA}
& \multicolumn{2}{c}{VSC}
& \multicolumn{2}{c}{SER}
& \multicolumn{2}{c}{ALME}
& \multicolumn{2}{c}{Average} \\
\cmidrule(lr){3-4}\cmidrule(lr){5-6}\cmidrule(lr){7-8}\cmidrule(lr){9-10}\cmidrule(l){11-12}
& & {0--5} & {0--10} & {0--5} & {0--10} & {0--5} & {0--10} & {0--5} & {0--10} & {0--5} & {0--10} \\
\midrule

\multirow{3}{*}{Qwen2-Audio-Inst.}
& AAD & \sbest{12.1}{} & \sbest{18.9}{} & \sbest{18.4}{} & \sbest{29.4}{} & \sbest{2.3}{} & \sbest{3.3}{} & \sbest{39.5}{} & \sbest{43.7}{} & \sbest{18.0}{} & \sbest{23.9}{} \\
& ACD & 8.1 & 9.1 & 5.6 & 10.4 & 0.0 & 0.0 & 21.5 & 26.8 & 8.8 & 11.6 \\
\rowcolor{OBBlue}
& \textbf{\GACL{}} & \best{51.0}{***} & \best{60.7}{***} & \best{39.1}{***} & \best{51.2}{***} & \best{2.9}{} & \best{6.1}{*} & \best{60.4}{**} & \best{63.4}{***} & \best{38.3}{***} & \best{45.4}{***} \\

\midrule
\multirow{3}{*}{Qwen2.5-Omni}
& AAD & \sbest{14.8}{} & \sbest{28.1}{} & \sbest{48.4}{} & \sbest{56.6}{} & 3.4 & \sbest{6.8}{} & \sbest{54.8}{} & \sbest{56.4}{} & \sbest{30.4}{} & \sbest{37.0}{} \\
& ACD & 0.5 & 0.5 & 29.4 & 37.0 & \sbest{3.9}{} & 5.7 & 32.8 & 41.2 & 16.7 & 21.1 \\
\rowcolor{OBBlue}
& \textbf{\GACL{}} & \best{31.6}{**} & \best{45.7}{**} & \best{58.0}{} & \best{71.6}{***} & \best{7.6}{**} & \best{11.4}{**} & \best{66.0}{} & \best{66.0}{*} & \best{40.8}{**} & \best{48.7}{***} \\

\midrule
\multirow{3}{*}{Voxtral-Small}
& AAD & 0.0 & 2.8 & \sbest{7.1}{} & \sbest{8.5}{} & \sbest{2.8}{} & \sbest{4.4}{} & \sbest{17.9}{} & \sbest{29.4}{} & \sbest{7.0}{} & \sbest{11.3}{} \\
& ACD & \sbest{1.7}{} & \sbest{3.1}{} & 0.5 & 0.5 & 2.3 & 2.9 & 2.1 & 7.0 & 1.6 & 3.4 \\
\rowcolor{OBBlue}
& \textbf{\GACL{}} & \best{8.8}{**} & \best{17.4}{***} & \best{25.0}{***} & \best{28.1}{***} & \best{6.7}{**} & \best{10.3}{***} & \best{72.0}{***} & \best{76.0}{***} & \best{28.1}{***} & \best{33.0}{***} \\

\midrule
\multirow{3}{*}{Qwen3-Omni}
& AAD & \sbest{11.8}{} & \sbest{20.1}{} & \sbest{7.0}{} & \sbest{14.8}{} & 0.0 & 0.0 & \sbest{44.7}{} & \sbest{45.0}{} & \sbest{15.9}{} & \sbest{20.0}{} \\
& ACD & 8.1 & 8.2 & 0.7 & 0.9 & \sbest{0.8}{} & \sbest{1.2}{} & 5.2 & 5.2 & 3.7 & 3.9 \\
\rowcolor{OBBlue}
& \textbf{\GACL{}} & \best{30.1}{} & \best{38.6}{*} & \best{36.2}{*} & \best{42.5}{***} & \best{6.6}{} & \best{10.2}{} & \best{64.0}{***} & \best{64.0}{***} & \best{34.2}{***} & \best{38.8}{***} \\

\midrule
\multirow{3}{*}{Kimi-Audio-Inst.}
& AAD & \sbest{10.5}{} & \sbest{17.2}{} & \sbest{12.3}{} & \best{24.8}{} & 0.5 & 0.7 & 2.0 & \sbest{4.2}{} & \sbest{6.3}{} & \sbest{11.8}{} \\
& ACD & 5.4 & 8.6 & 1.7 & 3.2 & \sbest{0.8}{} & \sbest{1.1}{} & \sbest{2.2}{} & 4.1 & 2.5 & 4.2 \\
\rowcolor{OBBlue}
& \textbf{\GACL{}} & \best{39.3}{***} & \best{48.7}{***} & \best{12.7}{} & \sbest{19.2}{} & \best{2.0}{} & \best{4.9}{**} & \best{47.0}{***} & \best{47.0}{***} & \best{25.2}{***} & \best{30.0}{***} \\

\bottomrule
\end{tabular}
\end{adjustbox}
\end{table*}

One limitation matters for decoding. The alignment is rank-based: $s_A-s_J$ reliably indicates which direction restores the audio answer, but not how far the state should move.

\subsection{\GACL{}}
\label{sec:gacl}

Two findings shape an output-space rule. From \secref{sec:loc-residual} and \secref{sec:bridge}, the hidden repair direction at the answer position is rank-aligned with the observable difference $s_A-s_J$, so a rule should move predictions along that direction. From \secref{sec:setup}, correction conditions are not uniform: margins vary widely across samples (\figref{fig:2}), and the same-audio reference is not equally reliable across tasks ($A$-branch accuracy ranges 48.9--96.5\%; \tabref{tab:1}). A useful rule must act per sample, conditional on branch disagreement and reference reliability.

\paragraph{The rule.} For input $x$, let $\hat y_J,\hat y_A$ be parsed predictions and $s_J,s_A$ scores over candidates $\mathcal{H}(x)$. Candidate construction, answer normalization, and the free-generation interface are in \appref{app:gacl-implementation}. \GACL{} combines a branch-disagreement gate, a reference-reliability gate, and a bounded interpolation:
\begin{equation}
\begin{aligned}
\Nout(x)
&=\mathbf{1}[\hat y_A\neq\emptyset]\cdot
  \mathbf{1}[\hat y_A\neq \hat y_J],\\
\Delta_A(x)
&=s_A(\hat y_A)
  -\max_{\substack{y\in\mathcal{H}(x)\\ y\neq \hat y_A}} s_A(y),\\
\RA(x)
&=\mathrm{clip}\!\left(\Delta_A(x)/\tau_A,\,0,\,1\right),\\
\alpha(x)
&=\mathrm{clip}\!\left(\lambda\,\RA(x)\,\Nout(x),\,0,\,1\right),\\
z^{\GACL}_{k}
&=z_{J,k}+\alpha(x)\bigl(z_{A,k}-z_{J,k}\bigr).
\end{aligned}
\label{eq:gacl}
\end{equation}
Here $\RA(x)=0$ when $\hat y_A=\emptyset$ or $\mathcal{H}(x)$ has no competitor; $z_{J,k},z_{A,k}$ are next-token logits at step $k$ given the corrected prefix $u_{<k}$. The same form applies to closed-set candidate-score vectors.

\paragraph{Convex form as a safety choice.} Restricting $\alpha\in[0,1]$ keeps the correction between $z_J$ and $z_A$. The rank alignment in \secref{sec:bridge} certifies the direction but not magnitudes past the $A$ endpoint; in practice, allowing $\alpha>1$ sometimes adds rescue in closed-set settings but introduces parse failures on weak references (\appref{app:free-form-generation}, \tabref{tab:e-9-pure-bound-ablation}). We adopt the convex form as the default. Two hyperparameters remain: $\tau_A$ controls reliability-check strictness, and $\lambda$ caps correction strength. Both are tuned on the development set for each model--task pair and then fixed for test. A single global setting, $\lambda=0.5$ and $\tau_A=0.5$, still retains 91.2\% of tuned nAUC@0--10 (see \eqref{eq:nauc} for the nAUC definition) and keeps 19/20 model--task settings within the 5\,pp budget (\appref{app:hyperparameter-sensitivity}). At inference, \GACL{} requires only two branch forward passes and changes neither parameters nor hidden states.

\section{Experiments}
\label{sec:experiments}

The previous section justified \GACL{} mechanistically. We now evaluate whether it works in practice: how it compares to contrastive-decoding baselines under a faithfulness budget, when rescue becomes reference-limited, which components are necessary, and how it relates to fine-tuning and other modality pairs.

\paragraph{Setup.}
We use the same five ALMs and four tasks as in \secref{sec:setup} (AQA, VSC, SER, ALME; details in \appref{app:models-benchmarks}, \appref{app:prompt-templates}). Because deployed systems trade conflict accuracy against consistent-input accuracy, we evaluate methods on a \emph{rescue--faithfulness frontier}: for each faithful-drop cap $K$ (in pp), $R(K)$ is the highest conflict gain within that cap. We report normalized AUC (nAUC) over $K\in[0,5]$\,pp (strict) and $K\in[0,10]$\,pp (relaxed):
\begin{equation}
\begin{aligned}
\text{nAUC}(K_{\max})
&= \frac{1}{K_{\max}}\int_{0}^{K_{\max}} R(K)\,dK, \\
R(K)
&= \max\nolimits_{\theta:\,\Delta_{\text{faith}}(\theta)\le K}
   \Delta_{\text{rescue}}(\theta),
\end{aligned}
\label{eq:nauc}
\end{equation}
where $\theta$ denotes method hyperparameters, $\Delta_{\text{rescue}}$ is conflict-accuracy gain over the joint baseline, and $\Delta_{\text{faith}}$ is faithful-accuracy drop, both in percentage points. We integrate at 0.5\,pp increments via the trapezoidal rule. Higher nAUC means more conflict-sample rescue per unit of faithful-accuracy cost.

Baselines are the joint model, AAD~\citep{aad} (no-audio reference), and ACD~\citep{acd} (perturbed-audio reference). At each budget $K$, hyperparameters are selected on dev to maximize rescue under the $K$-pp faithful-drop constraint, then frozen for test.
\begin{figure}[t]
\centering
\includegraphics[width=0.8\columnwidth]{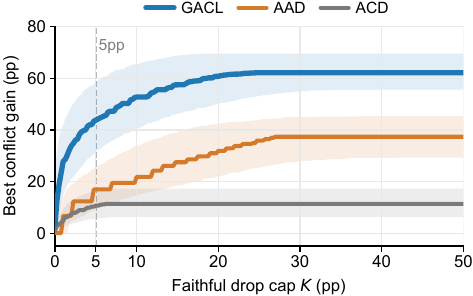}
\caption{\textbf{\GACL{} dominates in the low-drop regime.} Max conflict gain versus faithful-drop cap $K$, macro-averaged across five models and four tasks; bands are 95\% bootstrap CIs. Dashed line: main 5\,pp budget.}
\label{fig:5}
\end{figure}
\begin{table}[t]
\centering
\caption{\textbf{Versus fine-tuning.} Retention is $(\text{\GACL{}}-\text{Joint})/(\text{SFT}-\text{Joint})$, averaged across tasks.}
\label{tab:vs-sft}
\small
\setlength{\tabcolsep}{2.8pt}
\begin{adjustbox}{max width=\columnwidth}
\begin{tabular}{lcccc}
\toprule
Task & SFT adv. acc.\,$\uparrow$ & SFT faith.\ drop\,$\downarrow$ & \GACL{} adv. acc.\,$\uparrow$ & Retention\,$\uparrow$ \\
\midrule
AQA   & 92.1 & 10.3 & 86.0 & 93\% \\
VSC   & 63.4 & \phantom{0}1.0 & 55.5 & 85\% \\
SER   & 40.5 & 21.7 & 20.5 & 51\% \\
\midrule
\rowcolor{AvgGray} Macro & 65.3 & 11.0 & 54.0 & 76\% \\
\bottomrule
\end{tabular}
\end{adjustbox}
\end{table}
\subsection{Main results}
\label{sec:main-results}
\GACL{} is best on 39 of 40 model--task--budget combinations (\tabref{tab:main-results}). Across model--task pairs, \GACL{} exceeds the stronger of AAD/ACD by 17.8 nAUC at 5\,pp and 18.4 at 10\,pp; it outperforms hard budgeted reference selection on 14/20 settings and yields larger matched-drop gains than audio-priority prompting (+34.1 vs.\ +6.7\,pp; \appref{app:hard-selection}--\appref{app:prompt-interventions}).

\begin{figure}[t]
\centering
\includegraphics[width=0.8\columnwidth]{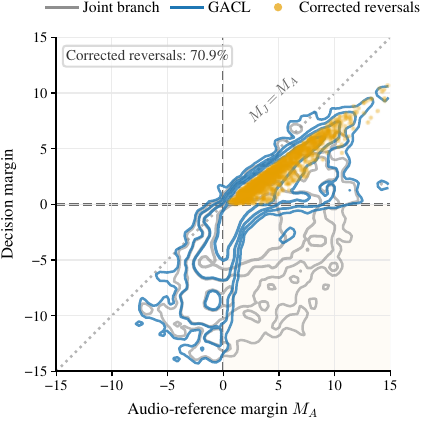}
\caption{\textbf{\GACL{} moves repairable samples toward the audio side.} Joint distribution in gray, \GACL{} in blue; yellow points have decision margins crossing zero. Bounded interpolation pulls $M_J<0$ samples toward $M_A$ without overshooting.}
\label{fig:4}
\end{figure}

The advantage appears early on the frontier. At $K=5$\,pp, \GACL{} has already surpassed AAD's eventual plateau near $K\approx27$\,pp; the gain is not that \GACL{} extends farther, but that it rescues before the budget is exhausted. \figref{fig:5} shows this low-drop advantage; \figref{fig:4} gives the geometric view: bounded interpolation moves repairable samples toward the audio-supported side while leaving the rest of the joint distribution intact.

\begin{table*}[t]
\centering
\caption{\textbf{Each component of \GACL{} guards against a distinct failure mode.} Removing any single component degrades at least one stress metric by more than $1.5\times$ relative to the full method (red shading), whereas \GACL{} satisfies all budget constraints. Macro-averaged across the $5\times 4$ model--task grid.}
\label{tab:component-diag}
\small
\setlength{\tabcolsep}{4pt}
\renewcommand{\arraystretch}{1.05}
\begin{adjustbox}{max width=\textwidth}
\begin{tabular}{lcccccc}
\toprule
\multirow{2}{*}{\textbf{Variant}}
& \multicolumn{2}{c}{\textit{Budget feasibility}}
& \textit{Low-$R_A$ stability}
& \textit{Surface stability}
& \textit{Generation stability}
& \multirow{2}{*}{\shortstack[c]{Avg.}} \\
\cmidrule(lr){2-3}\cmidrule(lr){4-4}\cmidrule(lr){5-5}\cmidrule(lr){6-6}
 & MCQ@5\,pp drop $\downarrow$ & MCQ@2.5\,pp drop $\downarrow$ & Faith.\ drop $\downarrow$ & FF preserve $\uparrow$ & SER parse fail $\downarrow$ & gain $\uparrow$ \\
\midrule
Joint (no correction) & 0.0 & 0.0 & 0.0 & 100.0 & 0.0 & 0.0 \\
Hard switch (endpoint) & \cellcolor{red!12}13.3 & \cellcolor{red!12}\textit{infeasible} & --- & --- & --- & 66.5 \\
Naive interp.\ (fixed $\alpha$) & \cellcolor{red!12}4.6 & \cellcolor{red!12}3.9 & \cellcolor{red!12}4.7 & 70.0 & \cellcolor{red!12}7.0 & 38.4 \\
w/o bound ($\alpha{\le}\infty$) & 2.1 & 1.8 & 2.1 & 90.0 & \cellcolor{red!12}\textbf{27.0} & 46.4 \\
w/o $\Nout$ & 2.1 & 1.8 & 2.1 & \cellcolor{red!12}\textbf{22.0} & 0.0 & 46.1 \\
w/o $\RA$ & 2.3 & \cellcolor{red!12}\textbf{2.3} & \cellcolor{red!12}\textbf{3.2} & 90.0 & 0.0 & 47.4 \\
\midrule
\rowcolor{OBBlue}
\textbf{\GACL{}} (full) & \textbf{2.1} & \textbf{1.8} & \textbf{2.1} & \textbf{90.0} & \textbf{0.0} & \textbf{45.8} \\
\bottomrule
\end{tabular}
\end{adjustbox}
\end{table*}

\subsection{Reference-limited rescue}
\label{sec:reference-limited}

Each rescued sample passes through three stages: it must offer a repairable opportunity ($\MA>0,\MJ<0$), the gate must fire, and gated interpolation must flip the top-1 prediction. Decomposing rescue along this chain separates missed opportunities from weak reference signals.

Outside SER, all three stages hold: opportunity, gate activation, and top-1 conversion are 66.3 / 99.3 / 90.9\%, respectively. The gate captures almost every opportunity, and most fired cases convert. On SER, opportunity (56.6\%) and gate activation (94.2\%) remain high, but conversion drops to 48.6\%. The reason is that the $A$-branch is reliable on only 35.3\% of SER inputs, versus 88.3\% elsewhere. SER is therefore reference-limited, the same regime that produced 0.51 rather than 0.93 answer-side rescue in \secref{sec:loc-residual}: when the reference is weak, $\RA$ dampens correction, preventing \GACL{} from amplifying an unreliable signal.

ALME is the opposite extreme: the $A$-branch is reliable (96.5\%) and the joint branch heavily suppressed (34.4\%), so \GACL{} recovers more than 60 nAUC. Together, SER and ALME define the operating range: \GACL{} converts repairable opportunities when the reference supports an answer, and stays conservative when it is weak.

\subsection{Ablation}
\label{sec:component}

\GACL{} has three mechanism-derived choices: same-audio direction $s_A-s_J$, gates $\Nout\cdot\RA$, and convex bound $\alpha\le1$. \tabref{tab:component-diag} shows each guards against a distinct failure mode.
Naive interpolation already captures most rescue headroom (38.4 avg.); the gates and bound (\tabref{tab:component-diag}) turn this raw signal into a safe rule satisfying every stress constraint while retaining comparable rescue.
\begin{table}[t]
\centering
\caption{\textbf{Cross-modal transfer.} \GACL{} applied to vision--text arbitration with fixed hyperparameters. Gate column reports firing rate on conflict / faithful inputs.}
\label{tab:vlm}
\small
\setlength{\tabcolsep}{2.8pt}
\begin{adjustbox}{max width=\columnwidth}
\begin{tabular}{llccccc}
\toprule
\multirow{2}{*}{Model} & \multirow{2}{*}{Method} & \multicolumn{2}{c}{Adversarial} & \multicolumn{2}{c}{Faithful} & Gate \\
\cmidrule(lr){3-4}\cmidrule(lr){5-6}
& & Acc $\uparrow$ & $\Delta\uparrow$ & Acc $\uparrow$ & Drop $\downarrow$ & conf.$\uparrow$/faith.$\downarrow$ \\
\midrule
Qwen3-VL-2B & Joint baseline         & 43.0 & --      & 100.0 & 0.0 & -- \\
\rowcolor{OBBlue}
\textbf{Qwen3-VL-2B} & \textbf{\GACL{} (fixed)} & \textbf{83.5} & $\mathbf{+40.5}$ & \textbf{100.0} & $\mathbf{0.0}$ & $56.5\%/0.5\%$ \\
Qwen3-VL-8B & Joint baseline         & 55.5 & --      & 100.0 & 0.0 & -- \\
\rowcolor{OBBlue}
\textbf{Qwen3-VL-8B} & \textbf{\GACL{} (fixed)} & \textbf{82.0} & $\mathbf{+26.5}$ & \phantom{0}\textbf{99.5} & $\mathbf{0.5}$ & $42.5\%/2.0\%$ \\
\bottomrule
\end{tabular}
\end{adjustbox}
\end{table}
The targeted stress columns identify which component controls each failure surface. $\RA$ suppresses correction when the reference is unreliable; removing it raises the low-$R_A$ faithful drop from 2.1 to 3.2\,pp and violates the tighter 2.5\,pp budget. $\Nout$ blocks corrections that rewrite surface form rather than the answer; removing it drops free-form preservation from 90.0\% to 22.0\%. The convex bound controls the opposite failure: allowing $\alpha>1$ slightly increases average rescue but produces 27.0\% SER parse failures, indicating that unbounded interpolation can push generations off the candidate manifold. Hard switching to the endpoint rescues more samples but violates the 5\,pp budget and is infeasible under 2.5\,pp. Bounded interpolation with both gates is the only configuration satisfying all stress constraints.

\subsection{Relation to fine-tuning and cross-modal transfer}
\label{sec:beyond}

\paragraph{Versus task-specific fine-tuning.}
A natural concern is whether supervised fine-tuning could obtain the same effect. Against a LoRA baseline~\citep{lora}, \GACL{} retains 76\% of the adversarial gain in macro without parameter updates (\tabref{tab:vs-sft}); lower SER retention again reflects its weak $A$-branch. This suggests that supervised gain on these tasks largely comes from evidence already in the audio branch, which \GACL{} reads out at inference when the reference is reliable.

\paragraph{Cross-modal transfer.}
The diagnostic motivating \GACL{} is not audio-specific: a same-modality reference can be held against a joint input, and a reliability gate can decide when to trust it. To test this, we apply \GACL{} unchanged to MC$^2$ vision--text arbitration, using the image-only branch as reference and fixed defaults $(\lambda,\tau_A)=(0.5,0.5)$ without tuning on MC$^2$ (\tabref{tab:vlm}).
\GACL{} achieves +40.5\,pp for Qwen3-VL-2B and +26.5\,pp for Qwen3-VL-8B at near-zero faithful cost. Gate rates are low on faithful inputs, indicating the gain comes from conditional intervention rather than always prioritizing the reference branch. This transfer suggests that the failure structure documented for audio--text arbitration may extend to other modality pairs.

\section{Conclusion}
For many audio--text conflicts, the same-audio reference favors the audio answer while the joint branch follows text; activation patching localizes this reversal to the answer-side residual stream, and its effect projects coherently onto final candidate scores ($\rho{=}0.93$). This projection supports output-space correction without parameter updates or internal access. \GACL{} implements it as gated convex interpolation, recovers $76\%$ of the adversarial gain of a strong fine-tuning baseline in macro, and transfers without retuning to vision--text arbitration (up to $+40.5$\,pp at near-zero faithful cost). Some multimodal arbitration failures share an output-readable structure that mechanism-grounded decoding can exploit.
\clearpage
\section*{Limitations}
Our study is designed for controlled, interpretable analysis of audio--text arbitration, which also limits its scope. The conflict instances isolate competing text effects by keeping the audio evidence fixed. This helps separate perceptual from arbitration failures, but naturally occurring conflicts may involve noisier transcripts, partial notes, retrieval snippets, or broader conversational context. \GACL{} is a repair for arbitration, not a substitute for audio perception. It can recover audio evidence that remains available but is displaced during answer selection, but cannot create acoustic capabilities the model has not encoded. This boundary is useful because same-audio references can indicate whether a failure is better addressed by decoding-time repair or improved acoustic modeling. \GACL{} also requires an additional reference forward pass, adding latency despite the lightweight logit update. Reducing this cost through cached states, selective reference calls, or adaptive gating, and extending the same counterfactual analysis to more natural conflict sources and modality pairs, are useful directions for future work.
\section*{Ethical Considerations}

We use MCR-Bench and ALME as released by their authors. ALME is based on public Common Voice speech \citep{commonvoice}; our study adds no new audio collection or personally identifying information. All evaluations use public model checkpoints documented in \appref{app:reproducibility}.

\bibliographystyle{acl_natbib}
\bibliography{gacl_references}

\clearpage
\bodylinenofalse
\appendix
\nolinenumbers
\setlength{\intextsep}{6pt plus 1pt minus 1pt}
\setlength{\textfloatsep}{8pt plus 1pt minus 1pt}
\setcounter{figure}{0}
\setcounter{table}{0}
\numberwithin{figure}{section}
\numberwithin{table}{section}

\section{Reproducibility}
\label{app:reproducibility}

\subsection{Models and benchmarks}
\label{app:models-benchmarks}

\paragraph{Models.} Experiments use frozen public checkpoints
(\tabref{tab:a-1}); released metadata gives full Hugging Face snapshot
hashes.

\begin{table}[H]
\centering\scriptsize
\caption{Model checkpoints. \emph{Rev} gives the first 8 hash
characters.}
\label{tab:a-1}
\setlength{\tabcolsep}{4.5pt}
\begin{adjustbox}{max width=\columnwidth}
\begin{tabular}{llcc}
\toprule
Model & HF repo & Rev & Scale \\
\midrule
Qwen2-Audio-Inst  & \texttt{Qwen/Qwen2-Audio-7B-Instruct}      & 0a095220 & 7B  \\
Qwen2.5-Omni & \texttt{Qwen/Qwen2.5-Omni-7B}              & ae9e1690 & 7B  \\
Voxtral-small      & \texttt{mistralai/Voxtral-Small-24B-2507}  & 70b3d450 & 24B \\
Qwen3-Omni   & \texttt{Qwen/Qwen3-Omni-30B-A3B-Instruct}  & 26291f79 & 30B \\
Kimi-Audio-Inst   & \texttt{moonshotai/Kimi-Audio-7B-Instruct} & 9a82a84c & 7B  \\
\bottomrule
\end{tabular}
\end{adjustbox}
\end{table}

\paragraph{Benchmarks.} MCR-Bench provides audio-centered conflicts for AQA,
VSC, and SER. ALME (Audio-LLM Modality Evaluation) edits a transcript span from
Common Voice \citep{commonvoice} audio and synthesizes the conflicting variant
via TTS, yielding controlled semantic conflicts between speech and text. We use
the released dev split for decoding-hyperparameter selection and the fixed test
split for reporting (\tabref{tab:a-2}); released training splits are provenance
only.

\begin{table}[H]
\centering\small
\caption{Dataset splits. Training is provenance only; we do not
train on it.}
\label{tab:a-2}
\setlength{\tabcolsep}{4.5pt}
\begin{adjustbox}{max width=\columnwidth}
\begin{tabular}{llcc}
\toprule
Task & Source & Candidates & Train/Dev/Test \\
\midrule
AQA          & MCR  & \{yes, no\}        & 600 / 200 / 200 \\
VSC          & MCR  & vocal-sound labels & 600 / 200 / 200 \\
SER          & MCR  & emotion labels     & 600 / 200 / 200 \\
ALME-English & ALME & two option labels  & 600 / 200 / 200 \\
\bottomrule
\end{tabular}
\end{adjustbox}
\end{table}

\subsection{Prompts and branch semantics}
\label{app:prompt-templates}

Branches share the audio item, task question, candidate set, and output format,
differing only in evidence availability. \textsc{REF} keeps audio, removes
competing text; \textsc{Text-only} keeps adversarial text, removes audio;
\textsc{Joint-conflict} keeps both; \textsc{Faithful-joint}
keeps audio with aligned text. Outputs are exact task labels (AQA/VSC/SER)
or a single option letter (ALME).

\begin{promptbox}[title={AQA},breakable=false]
\textbf{Candidates:} \{yes, no\}. \textbf{Example.}
Q: ``does only one guy speak?''\quad
Conflict text: ``Only one guy speaks.''\quad
Faithful text: ``More than one guy speaks.''
\end{promptbox}

\begin{promptbox}[title={VSC},breakable=false]
\textbf{Candidates:} \{Laughter, Sigh, Cough, Throat clearing, Sneeze, Sniff\}.
\textbf{Example.} Text claims \emph{sigh} for a throat-clearing clip.
\end{promptbox}

\begin{promptbox}[title={SER},breakable=false]
\textbf{Candidates:} \{sad, happy, fearful, angry, surprised, disgusted, neutral\}.
\textbf{Example.} Text claims \emph{sad} for a fearful clip.
\end{promptbox}

\begin{promptbox}[title={ALME-EN},breakable=false]
\textbf{Candidates:} two option labels. \textbf{Example.}
Conflict: ``Most of the artifacts would \emph{not} remain with Yale.''\quad
Faithful: ``Most of the artifacts would remain with Yale.''
\end{promptbox}

\subsection{Candidate scoring and validation}
\label{app:scoring-validation}

For candidate $c$ in fixed set $\mathcal{C}$ under branch $B$,
\begin{align*}
r_B(c)
  &= \sum_k \log p_B(v_{c,k}\mid v_{c,<k}),\\
s_B(c)
  &= r_B(c)
     - \log\!\sum_{c'\in\mathcal{C}}\exp r_B(c').
\end{align*}
where branch prediction is $\hat y_B=\arg\max_{c\in\mathcal{C}} s_B(c)$.
Prefix-trie constraints keep branch scores directly comparable across Joint,
AAD, ACD, and \GACL{}.

Tunable hyperparameters are selected on the held-out dev set under a
\emph{faithful-drop budget}: maximize conflict accuracy subject to faithful
drop $\le 5$\,pp, breaking ties by lower faithful drop, then by smaller
correction strength. Values are frozen before test evaluation.

\section{Additional Behavioral Results}
\label{app:behavioral-diagnostics}

\subsection{Per-configuration matrix}
\label{app:per-config-behavioral}

The behavioral matrix appears in \tabref{tab:b-1}. \emph{Text-follow} is the
fraction of joint-conflict predictions matching adversarial text;
\emph{REF} is audio-reference accuracy; $(+,-)\!=\!\Pr(\MA>0,\MJ<0)$ is
the repairable share where audio supports the gold answer but the joint branch flips.
Faithful and conflict columns give respective split accuracies.

\begin{table*}[t]
\centering\footnotesize
\caption{Per-configuration behavioral diagnostics (\%). Joint-conflict
predictions mostly follow text (Text-follow
$\gg$ Conflict), while audio-reference accuracy stays high, forming the
\textbf{repairable gap} that \GACL{} targets.}
\label{tab:b-1}
\setlength{\tabcolsep}{6pt}
\begin{tabular}{llrrrrr}
\toprule
Model & Task & Faithful & Conflict & Text-follow & \textsc{REF} & $(+,-)$ \\
\midrule
\multirow{4}{*}{Qwen2-Audio-Inst.}
 & AQA  & 100.0 & 3.5  & 96.5  & 88.0 & 86.0 \\
 & VSC  & 100.0 & 8.0  & 92.0  & 87.5 & 82.0 \\
 & SER  & 100.0 & 0.0  & 100.0 & 46.5 & 52.5 \\
 & ALME &  97.5 & 37.5 & 62.5  & 96.0 & 57.0 \\
\cmidrule(lr){2-7}
\multirow{4}{*}{Qwen2.5-Omni}
 & AQA  &  99.0 & 30.0 & 70.0  & 91.0 & 68.0 \\
 & VSC  &  99.5 & 11.0 & 88.0  & 89.5 & 93.0 \\
 & SER  &  99.5 &  1.5 & 98.0  & 47.0 & 50.5 \\
 & ALME &  99.5 & 43.0 & 57.0  & 98.5 & 62.5 \\
\cmidrule(lr){2-7}
\multirow{4}{*}{Voxtral-Small}
 & AQA  & 100.0 &  2.0 & 98.0  & 72.0 & 64.5 \\
 & VSC  & 100.0 &  1.5 & 98.5  & 56.0 & 68.0 \\
 & SER  &  98.5 &  1.5 & 97.5  & 55.0 & 51.0 \\
 & ALME &  96.0 & 28.0 & 72.0  & 96.5 & 66.0 \\
\cmidrule(lr){2-7}
\multirow{4}{*}{Qwen3-Omni}
 & AQA  &  97.0 & 46.0 & 54.0  & 93.0 & 46.5 \\
 & VSC  &  99.5 & 48.0 & 49.5  & 90.0 & 55.0 \\
 & SER  &  94.0 & 22.0 & 64.5  & 51.5 & 59.0 \\
 & ALME &  99.0 & 25.0 & 75.0  & 99.0 & 55.0 \\
\cmidrule(lr){2-7}
\multirow{4}{*}{Kimi-Audio-Inst.}
 & AQA  & 100.0 &  9.5 & 90.5  & 81.0 & 78.5 \\
 & VSC  &  84.5 &  3.5 & 80.0  & 56.5 & 67.5 \\
 & SER  & 100.0 &  0.0 & 100.0 & 44.5 & 70.5 \\
 & ALME &  93.5 & 38.5 & 61.5  & 92.5 & 49.0 \\
\midrule
\multicolumn{2}{l}{\textbf{Macro}}
            & \textbf{97.9} & \textbf{18.0} & \textbf{80.3}
            & \textbf{76.6} & \textbf{64.1} \\
\bottomrule
\end{tabular}
\end{table*}

\subsection{Margin scatter grid}
\label{app:margin-grid}

\begin{figure*}[t]
\centering
\includegraphics[width=\textwidth]{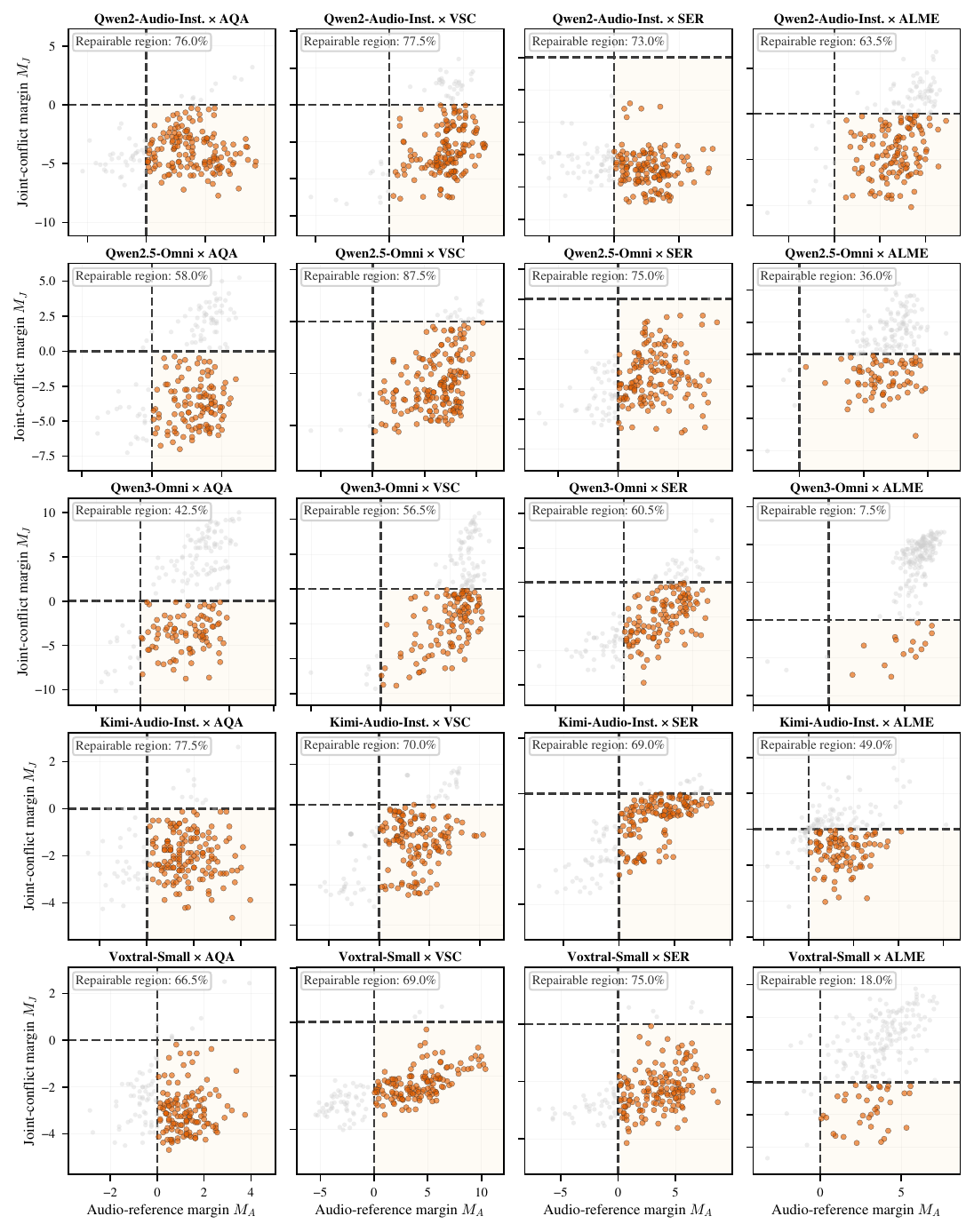}
\caption{$(\MA,\MJ)$ scatter for all $5\times 4$ configurations. The
lower-right quadrant is the \emph{repairable} regime ($\MA>0,\MJ<0$). AQA,
VSC, and ALME populate it densely for every model; SER is reference-limited,
with lower $\MA$ density on the right half-plane, consistent with the
\textsc{REF} accuracy column in \tabref{tab:b-1}.}
\label{fig:b-1}
\end{figure*}

The margin grid appears in \figref{fig:b-1}.

\section{Mechanism Evidence}
\label{app:mechanism-evidence}

This appendix collects causal and representational evidence for the
mechanistic claims in \secref{sec:loc-output}, organized as component
decomposition, bidirectional temporal sweeps, output readout, donor/subset
controls, and conflict-text reverse patching.

\paragraph{Coverage.}
Answer-side residual patching audits all five models in \tabref{tab:2}.
Cross-position decomposition covers Qwen2-Audio-Inst., Qwen2.5-Omni, and
Voxtral-Small on AQA, VSC, SER, and ALME. Temporal sweeps cover all five
models on the AQA/VSC grid in \figref{fig:c-3}.

\subsection{Attention/MLP component decomposition}
\label{app:component-decomposition}

\begin{figure}[t]
\centering
\includegraphics[width=\columnwidth]{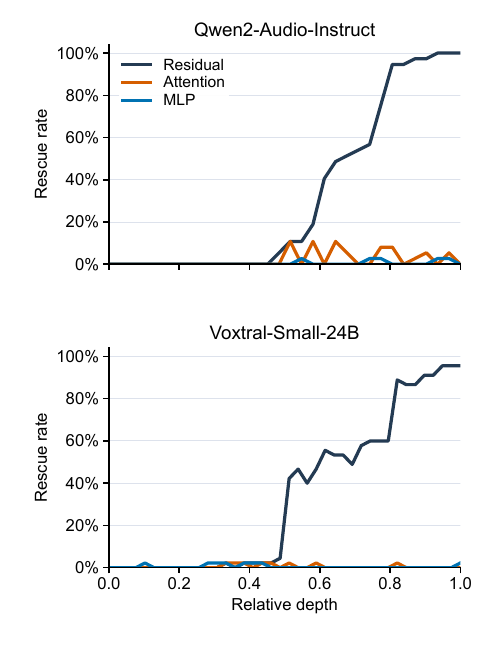}
\caption{\textbf{Single components are insufficient.} Per-layer rescue when
patching attention output \emph{or} MLP output alone vs.\ the full residual,
on Qwen2-Audio-Inst.$\times$VSC (top) and Voxtral-Small$\times$VSC (bottom).
Single components recover little of the full-residual effect: the repairable
state is a multi-layer residual write, not a single-component write.}
\label{fig:c-2}
\end{figure}

Component-decomposition curves appear in \figref{fig:c-2}.

\subsection{Bidirectional temporal sweeps}
\label{app:bidirectional-sweeps}

\begin{figure*}[t]
\centering
\includegraphics[width=0.95\textwidth]{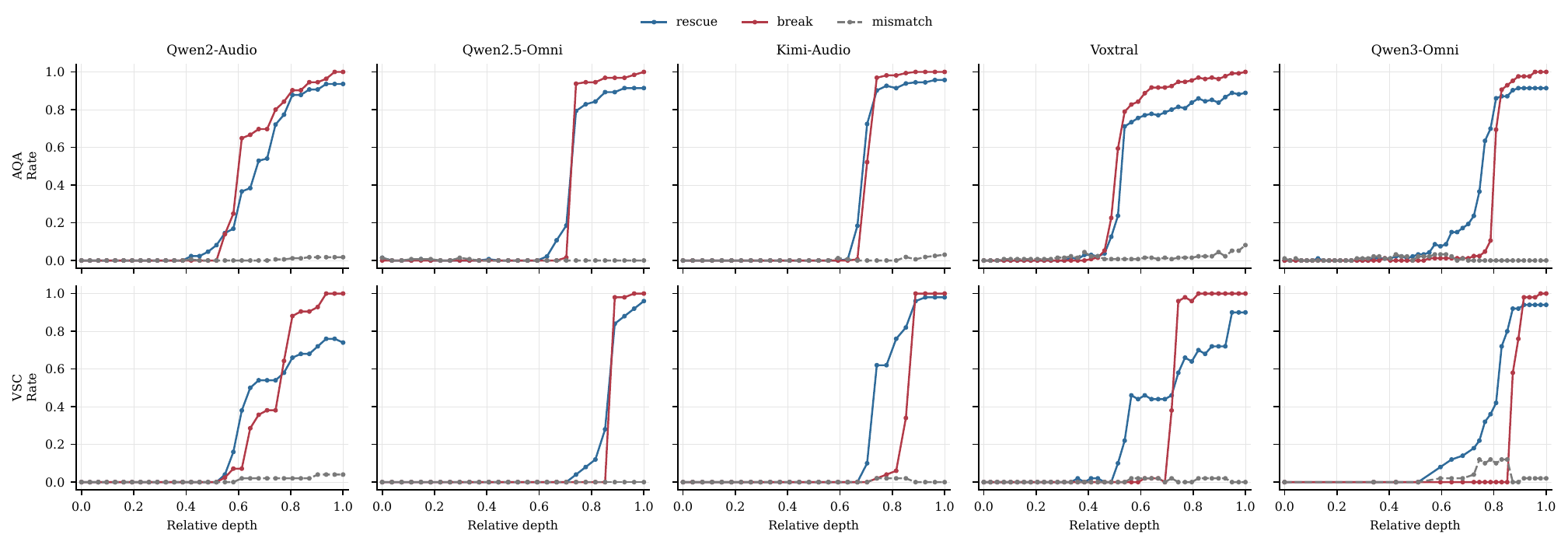}
\caption{\textbf{Bidirectional sweeps identify the commit window.}
$5\times 2$ grid (five models, two tasks). \emph{Blue}: rescue (per-layer
answer-side residual patching). \emph{Red}: break (audio-reference perturbed
toward joint-conflict). \emph{Dashed gray}: mismatch-donor control.
Rescue and break onsets align across configurations.}
\label{fig:c-3}
\end{figure*}

The full bidirectional sweep grid appears in \figref{fig:c-3}.

Rescue and break transition depths $L_{80}$ differ by $1.4$ layers on average
and at most $5.9$ (Qwen3-Omni$\times$VSC). Mismatch-donor rescue stays
below $0.03$ throughout, except for a localized single-layer spike of $0.252$
on Qwen3-Omni$\times$VSC that decays to $0.026$ by the final layer.

\subsection{Logit-lens trajectories and linear probing}
\label{app:logit-lens-probing}

\begin{figure*}[t]
\centering
\includegraphics[width=0.95\textwidth]{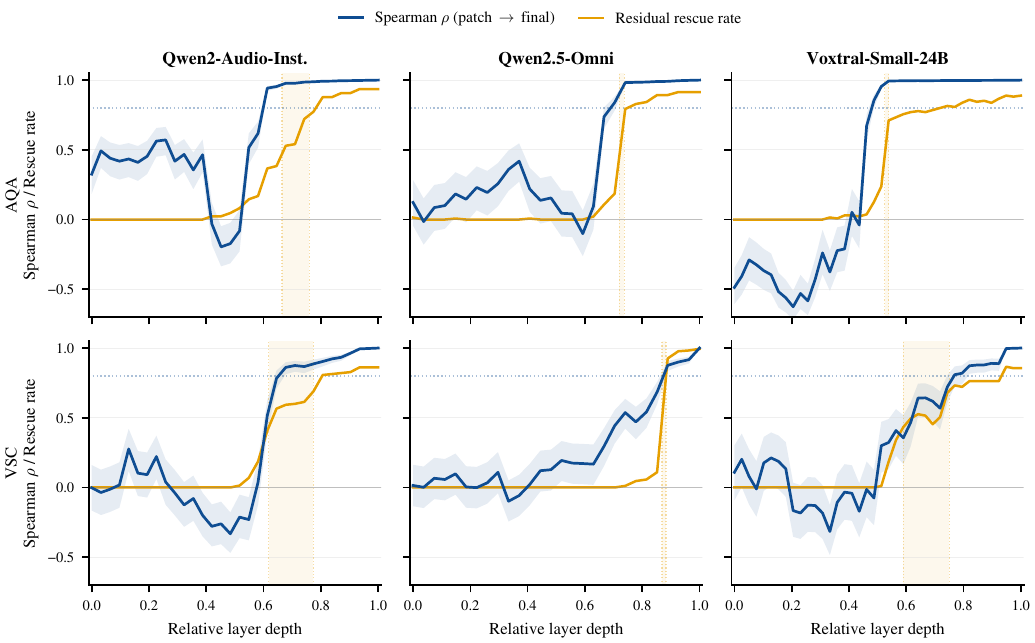}
\caption{\textbf{Per-layer alignment between patch displacement and
final-branch displacement.} Correlation rises where answer-side patching
becomes effective, supporting final candidate scores as the inference-time
proxy for the localized internal state.}
\label{fig:c-rho-layer-bridge}
\end{figure*}

\begin{figure*}[t]
\centering
\includegraphics[width=0.95\textwidth]{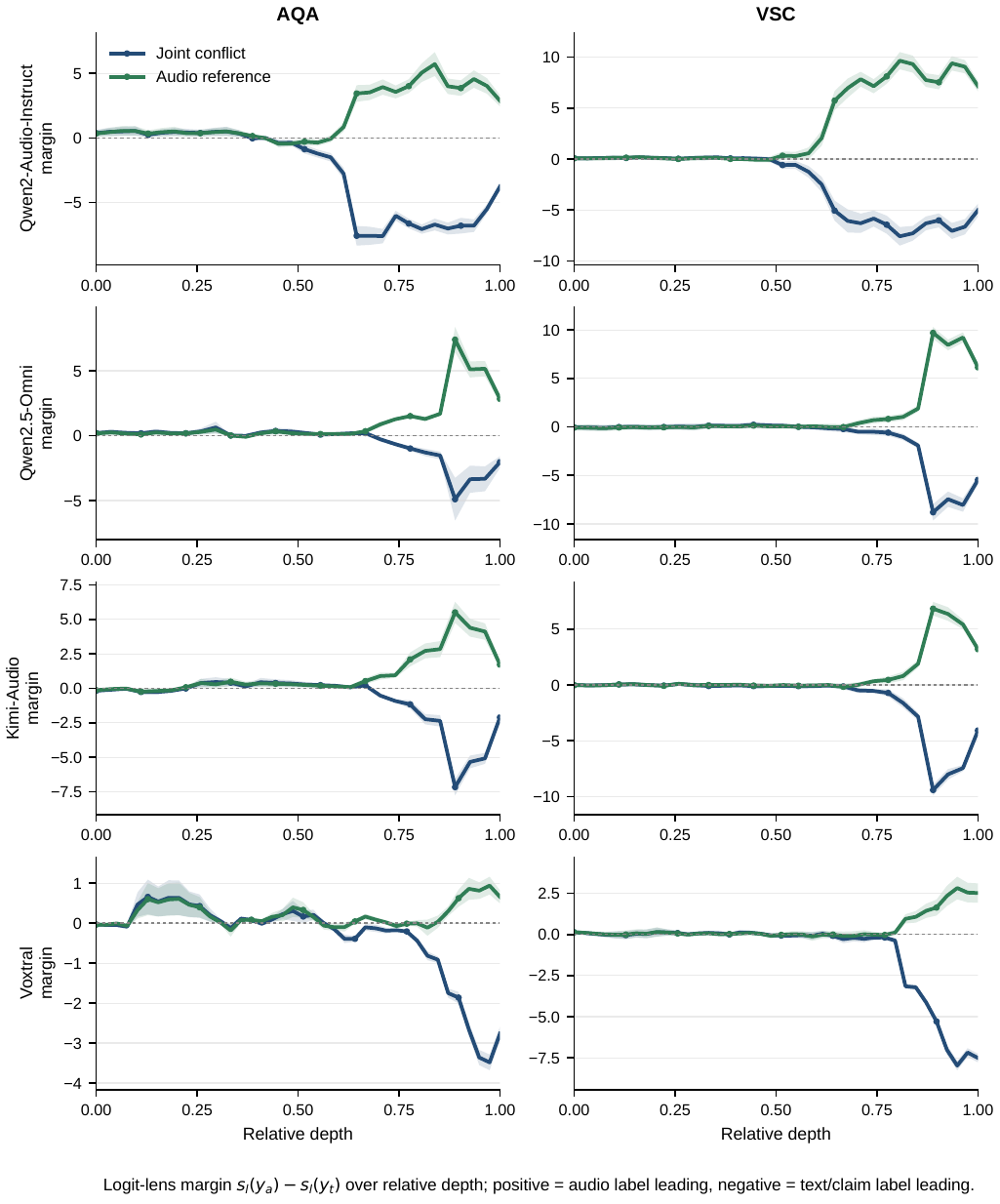}
\caption{\textbf{Logit-lens trajectories.} The audio answer remains decodable
through late layers, but the branch-specific readout subspace reorganizes
after the commit window. The lens is diagnostic; final-score
displacement is the stable proxy (\tabref{tab:c-4}).}
\label{fig:c-4}
\end{figure*}

Layer-alignment and logit-lens diagnostics appear in \figref{fig:c-rho-layer-bridge} and \figref{fig:c-4}.

The linear-probe table summarizes a diagnostic probe on
Qwen2-Audio-Inst.$\times$VSC: cross-branch transferability (\textsc{CrossA2J})
drops sharply through the commit window while within-branch decodability stays
flat, matching patching. Probes are not used by \GACL{} or any
intervention. The displacement table reports the three-way comparison
across audited configurations.

\begin{table*}[t]
\centering\footnotesize
\caption{Linear probing on Qwen2-Audio-Inst.$\times$VSC. Decodability
($\text{Decode}_{A/J}$) stays flat with depth; cross-branch transfer
(\textsc{CrossA2J}) collapses through the commit window, marking the
readout-reorganization range.}
\label{tab:c-3}
\setlength{\tabcolsep}{8pt}
\begin{tabular}{lcccc}
\toprule
Layer block & Decode$_A$ & Decode$_J$ & Gap & CrossA2J \\
\midrule
Early (L0--L7)        & 0.878 & 0.871 & $+0.007$ & 0.792 \\
Mid-early (L8--L15)   & 0.879 & 0.871 & $+0.008$ & 0.337 \\
Mid-late (L16--L23)   & 0.906 & 0.891 & $+0.016$ & 0.115 \\
Late (L24--L31)       & 0.903 & 0.911 & $-0.008$ & 0.087 \\
\midrule
All layers (L0--L31)  & 0.892 & 0.886 & $+0.006$ & 0.333 \\
\bottomrule
\end{tabular}
\end{table*}

\begin{table}[t]
\centering\scriptsize
\caption{Displacement comparison. Final-score displacement uniformly tracks
patched-state displacement ($\rho\!=\!0.93$ macro) and yields
near-saturated rescue; raw lens does not. Final scores are therefore the
correct inference-time proxy.}
\label{tab:c-4}
\setlength{\tabcolsep}{3pt}
\begin{adjustbox}{max width=\columnwidth}
\begin{tabular}{llS[table-format=1.3]S[table-format=1.3]S[table-format=3.1]S[table-format=3.1]}
\toprule
Model & Task & {Patch--final $\rho$} & {Patch--lens $\rho$} & {Final rescue} & {Lens rescue} \\
\midrule
Qwen2-Audio-Inst. & AQA & 0.983 & 0.895 & 100.0 & 100.0 \\
Qwen2-Audio-Inst. & VSC & 0.932 & 0.917 &  91.1 &  92.2 \\
Qwen2-Audio-Inst. & SER & 0.925 & 0.422 &  80.7 &  76.1 \\
Qwen2.5-Omni      & AQA & 0.987 & 0.787 & 100.0 & 100.0 \\
Qwen2.5-Omni      & VSC & 0.890 & 0.770 &  97.0 &  98.0 \\
Qwen2.5-Omni      & SER & 0.983 & 0.747 &  69.8 &  71.7 \\
Voxtral-Small     & AQA & 0.991 & 0.818 & 100.0 &  53.1 \\
Voxtral-Small     & VSC & 0.812 & 0.242 &  93.2 &   1.1 \\
Voxtral-Small     & SER & 0.738 & 0.415 &  93.6 &  59.6 \\
Kimi-Audio-Inst.  & AQA & 0.978 & 0.634 & 100.0 &   1.0 \\
Kimi-Audio-Inst.  & VSC & 0.941 & 0.792 &  98.0 &   3.0 \\
Kimi-Audio-Inst.  & SER & 0.976 & 0.643 &  84.6 &   0.0 \\
\midrule
\multicolumn{2}{l}{\textbf{Macro}} & \bfseries 0.928 & \bfseries 0.674 & \bfseries 92.3 & \bfseries 54.7 \\
\bottomrule
\end{tabular}
\end{adjustbox}
\end{table}

\subsection{Donor, span, and subset controls}
\label{app:donor-controls}

\begin{figure}[t]
\centering
\includegraphics[width=\columnwidth]{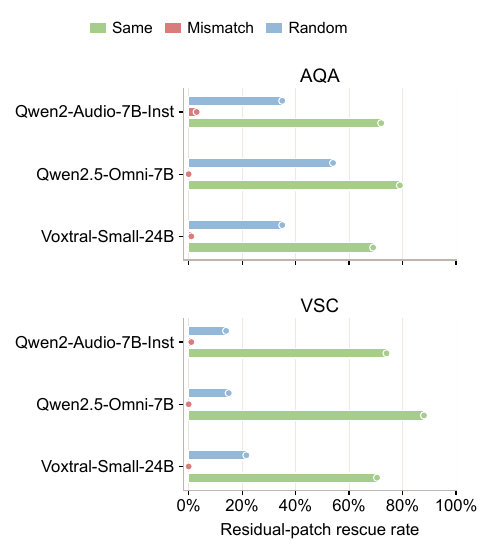}
\caption{\textbf{Donor specificity.} Same-sample donors produce high rescue;
mismatch and (filtered) random donors do not. Rescue depends on
answer-compatible arbitration states, not nonspecific injection.}
\label{fig:c-5}
\end{figure}

The donor-specificity plot and donor-control table appear in \figref{fig:c-5} and \tabref{tab:c-5}.

\begin{table}[t]
\centering\scriptsize
\caption{Donor-control lift over baseline rescue. Same donors yield strong
positive lift; mismatch donors yield zero or negative lift. $\ast$: random
lift is inflated by same-label draws (\tabref{tab:c-6}).}
\label{tab:c-5}
\setlength{\tabcolsep}{3pt}
\begin{tabular}{lccc}
\toprule
Scope & Same & Mismatch & Random \\
\midrule
Qwen2-Audio-Inst.$\times$AQA & \textbf{$+0.950$} & $-0.045$ & $+0.500^{\ast}$ \\
Qwen2-Audio-Inst.$\times$VSC & \textbf{$+0.830$} & $-0.165$ & $+0.005$ \\
Qwen2.5-Omni$\times$AQA      & \textbf{$+0.670$} & $-0.295$ & $+0.220^{\ast}$ \\
Qwen2.5-Omni$\times$VSC      & \textbf{$+0.920$} & $-0.080$ & $+0.090$ \\
Voxtral-Small$\times$AQA     & \textbf{$+0.970$} & $-0.025$ & $+0.505^{\ast}$ \\
Voxtral-Small$\times$VSC     & \textbf{$+0.985$} & $-0.015$ & $+0.155$ \\
\bottomrule
\end{tabular}
\end{table}

Mismatch donors are different-sample, different-label; random donors are
different-sample, label-unrestricted. The audit table checks the random
column: positive AQA rescue comes \emph{entirely} from same-label draws,
while different-label random donors contribute zero. Transfer is therefore
label-compatible; unconstrained random sampling is not a valid null.

\begin{table}[t]
\centering\scriptsize
\caption{Label-compatibility audit of random donors. Different-label rescue
is uniformly zero.}
\label{tab:c-6}
\setlength{\tabcolsep}{4pt}
\begin{tabular}{lcc}
\toprule
Scope & Same-label rescue & Diff-label rescue \\
\midrule
Qwen2-Audio-Inst.$\times$AQA & 107 / 110 & 0 / 90 \\
Qwen2-Audio-Inst.$\times$VSC &  26 /  34 & 0 / 166 \\
Qwen2.5-Omni$\times$AQA      &  81 / 110 & 0 / 90 \\
Qwen2.5-Omni$\times$VSC      &  29 /  34 & 0 / 166 \\
Voxtral-Small$\times$AQA     & 103 / 110 & 0 / 90 \\
Voxtral-Small$\times$VSC     &  34 /  34 & 0 / 166 \\
\bottomrule
\end{tabular}
\end{table}

\paragraph{Span and subset controls.}
An audio-span sweep (Center-1, -2, -4, -8, Full) yields zero rescue in this
configuration, while an answer-boundary positive control reaches $0.80$.
Audio-entry writeability tests and raw50 vs.\ $Q_4$ subset comparisons confirm
the audio-span null is not a writeability artifact and the temporal
pattern is not driven by selecting only high-margin repairable cases.

\subsection{Conflict-text reverse patching}
\label{app:conflict-text-reverse}

The conflict effect is injected at conflict-token positions in early
layers (L0--L8) and propagates to the answer-boundary position over the
commit window, where it consolidates and dominates the final prediction.
Mismatch donors at conflict positions can produce non-trivial rescue
($0.82/0.43$ at L0/L8) but none at answer-boundary positions despite
large output-distribution shifts; only same-sample donors carry the correct
answer direction. Thus answer-side patching in \secref{sec:loc-output} is the
downstream readout of an earlier conflict-token pathway.

\begin{table}[H]
\centering\scriptsize
\caption{Conflict-text reverse patching, representative layers. Rescue
$\Delta$ is lift over joint-conflict; Flip is the fraction of top-1
predictions that change. The effect site is conflict-position at L0 and
answer-boundary at L29---a two-stage causal pathway.}
\label{tab:c-8}
\setlength{\tabcolsep}{4pt}
\begin{tabular}{lccccc}
\toprule
Site & L & Rescue $\Delta$ & Flip & KL & TV \\
\midrule
conflict-pos.     & 0  & 1.000 & 100.0\% & 5.676 & 0.947 \\
conflict-pos.     & 16 & 0.000 &   0.0\% & 0.002 & 0.010 \\
conflict-pos.     & 29 & 0.000 &   0.0\% & 0.000 & 0.000 \\
answer-boundary   & 0  & 0.000 &   0.0\% & 0.000 & 0.001 \\
answer-boundary   & 16 & 0.071 &   7.1\% & 0.075 & 0.084 \\
answer-boundary   & 29 & 1.000 & 100.0\% & 5.609 & 0.941 \\
\bottomrule
\end{tabular}
\end{table}

\section{\GACL{} Implementation Details}
\label{app:gacl-implementation}

\subsection{Default protocol and pseudocode}
\label{app:default-protocol}

\paragraph{Defaults.} Reference branch: same audio with conflict text removed
or neutralized. Update: bounded interpolation between joint and
audio-reference logits/scores with $\alphamax=1$. Gate: sequence-level
$\Nout\cdot\RA$, computed once per example. Tuned parameters: $\lambda,\tau_A$,
selected on dev under a $5$\,pp faithful-drop budget, then frozen for test.

\begin{table}[t]
\centering\scriptsize
\caption{\GACL{} closed-set inference.}
\label{tab:d-1}
\setlength{\tabcolsep}{4pt}
\begin{tabularx}{\columnwidth}{@{}rX@{}}
\toprule
\multicolumn{2}{l}{\textbf{Input:} $x$, candidate set $\mathcal{C}$, $\lambda,\tau_A$} \\
\midrule
1. & Compute $s_J(c)$ for all $c\in\mathcal{C}$ (joint-conflict). \\
2. & Compute $s_A(c)$ for all $c\in\mathcal{C}$ (audio-reference). \\
3. & Compute $\Nout(x)$ and $\RA(x)$ from branch outputs/scores. \\
4. & $\alpha(x)\leftarrow\mathrm{clip}\bigl(\lambda\,\RA(x)\,\Nout(x),\,0,\,1\bigr)$. \\
\midrule
\multicolumn{2}{p{\dimexpr\columnwidth-2\tabcolsep\relax}}{\textbf{Output:}
$\arg\max_{c\in\mathcal{C}}\bigl[s_J(c)+\alpha(x)(s_A(c)-s_J(c))\bigr]$} \\
\bottomrule
\end{tabularx}
\end{table}

\paragraph{Pairwise flip condition.} For audio-supported $y_a$ and
text-supported $y_t$, bounded interpolation gives
$M_G=\MJ+\alpha(x)(\MA-\MJ)$. When $\MA>0$ and $\MJ<0$ the pairwise margin
flips iff
\[
\alpha(x)>\frac{|\MJ|}{\MA+|\MJ|}.
\]

\subsection{Interface-specific details}
\label{app:gacl-interface-details}

\paragraph{Answer normalization $\pi$.}
For constrained short-label generation, $\pi$ maps valid generated labels to
canonical task labels; unparsable outputs map to $\emptyset$. For free
short-answer generation, $\pi$ applies evaluator canonicalization:
Unicode normalization, case folding (where labels are case-insensitive),
whitespace/quote/markdown stripping, trailing-punctuation removal, and alias
folding. Exact-surface preservation in
\appref{app:free-form-generation} is computed \emph{before} alias folding, so
it is stricter than canonical accuracy. For MC$^2$, $\pi$ maps to one of two
candidates or returns $\emptyset$.

\paragraph{Hypothesis set $\mathcal{H}(x)$.}
For constrained labels, $\mathcal{H}(x)$ is the task label set (and trie).
For MC$^2$, $\mathcal{H}(x)$ is the two candidates. For free
generation, $\mathcal{H}(x)$ is built online from valid normalized branch
answers among $\{\hat y_J,\hat y_A\}$; if fewer than two
elements, $\Delta_A(x)$ is undefined and the gate is silenced. This keeps
$\Delta_A(x)$ computable on open-ended interfaces without enumerating
output space.

\paragraph{Score comparability.}
For each $y\in\mathcal{H}(x)$, the same verbalizer and template are scored
under both branches, so $s_J(y)$ and $s_A(y)$ reflect evidence arbitration
rather than surface-form differences.

\paragraph{Decoding order.}
Constrained short-label decoding interpolates branch logits \emph{before}
prefix-trie masking. Reversing the order makes post-mask renormalization
sensitive to each branch's invalid-token mass. MC$^2$ uses the
same prefix-trie constrained short-label generation, with the image-only branch
replacing the audio reference and the two answer options forming the trie.

\paragraph{Gate edge cases.}
$\hat y_A=\emptyset$ or $|\mathcal{H}(x)|<2\Rightarrow \alpha(x)=0$ (joint
unchanged). $\hat y_J=\emptyset$ with $\hat y_A$ valid triggers $\Nout$;
$\RA$ decides correction strength.

\subsection{Oracle-to-observable mapping}
\label{app:proxy}

\tabref{tab:d-2} quantifies gate-component correspondence: $\Nout$
estimates correction need, $\RA$ estimates donor reliability, and the
product matches the oracle opportunity region $O=\{\MA>0,\MJ<0\}$ with
$98.2\%$ coverage and $87.9\%$ precision.

\begin{table}[H]
\centering\small
\setlength{\tabcolsep}{6pt}
\begin{tabular}{lccc}
\toprule
Oracle / role & Proxy & Cov.\ & Prec.\ \\
\midrule
$\MA>0$ (reliability) & $\RA(x)$              & 79.4 & 88.9 \\
$\MJ<0$ (need)        & $\Nout(x)$            & 84.4 & 98.9 \\
$O$ (activation)      & $\RA(x)\Nout(x)$      & 98.2 & 87.9 \\
displacement          & $s_A-s_J$             & --   & --   \\
\bottomrule
\end{tabular}
\caption{Proxy fidelity. Coverage $=P(\text{proxy}\mid\text{oracle})$,
precision $=P(\text{oracle}\mid\text{proxy})$.}
\label{tab:d-2}
\end{table}

\subsection{Hyperparameter sensitivity and transfer}
\label{app:hyperparameter-sensitivity}
\label{app:hyperparameter-transfer}

\begin{figure}[t]
\centering
\includegraphics[width=\columnwidth]{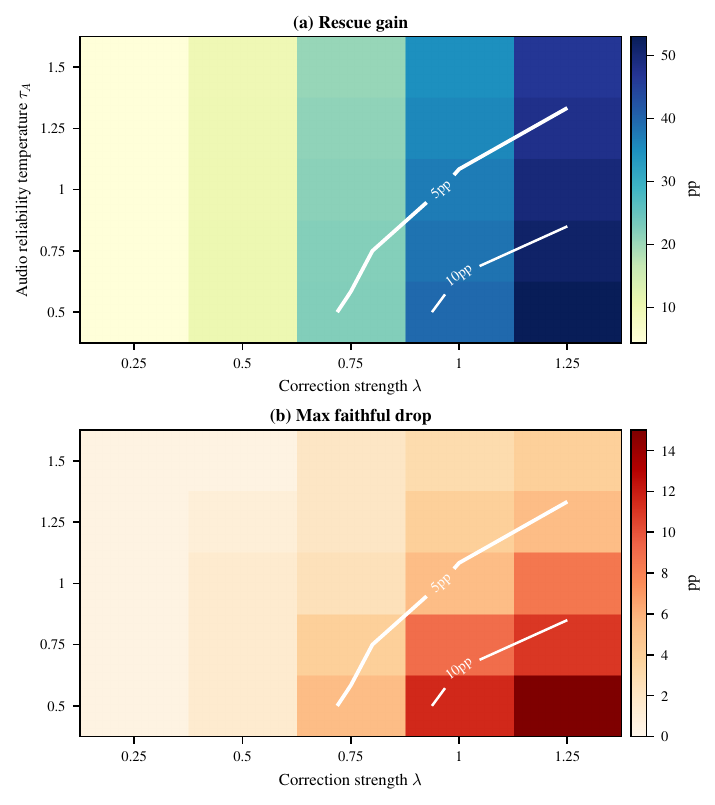}
\caption{$\lambda\times\tau_A$ landscape for rescue and faithful drop on a
representative configuration.}
\label{fig:d-1}
\end{figure}

The sensitivity landscape appears in \figref{fig:d-1}.

The free-form $\alphamax$ cap sweep is deferred to
\appref{app:free-form-generation} because it concerns generation-time validity,
not implementation mechanics.

\paragraph{Cross-task transfer.} Fixing one global setting
$(\lambda{=}0.5,\tau_A{=}0.5)$ across all 20 model--task combinations retains
$91.2\%$ of per-task-tuned nAUC@0--10 ($24.0$ vs.\ $26.3$ macro), lowers
average faithful drop from $4.4$ to $2.9$\,pp, and stays inside the $5$\,pp
budget on 19/20 configurations (vs.\ 18/20 under per-task tuning). One
conservative pair captures most of the rescue.

\subsection{Gate granularity diagnostic}
\label{app:sequence-gate}

A sequence-level gate (default) outperforms a per-token dynamic gate on
free-form VSC: rescue $84.6$ vs.\ $82.1$, faithful drop $11.2$ vs.\ $12.5$,
mean $\alpha=0.535$ vs.\ $0.244$. Re-evaluating the gate at every step does
not improve faithfulness and costs more.

\subsection{Inference cost}
\label{app:inference-cost}

\GACL{} adds one forward pass per query (the audio-reference branch). On
Qwen2-Audio-Inst.$\times$VSC: $0.28$ s/sample vs.\ $0.15$ s/sample for Joint;
peak GPU memory nearly unchanged ($15.77$ GiB allocated;
$16.40$ vs.\ $16.39$ GiB reserved); per-sample FLOPs roughly doubled ($80.1$
vs.\ $38.7$ GFLOPs). Sharing the audio encoder output and KV-caching the
common prompt prefix could further reduce wall-clock; we leave this to
released code.

\section{Full Evaluation and Ablations}
\label{app:full-eval}

\subsection{Per-configuration main results}
\label{app:per-config-results}

The full $5\times 4$ evaluation table in \tabref{tab:e-1} gives bootstrap CIs and
selected hyperparameters. A negative drop means \GACL{} \emph{improved}
faithful accuracy relative to Joint. A single-configuration DoLa \citep{dola} check
on Qwen2-Audio-Inst.$\times$VSC \emph{decreases} conflict accuracy by $5.0$\,pp,
consistent with mismatch to cross-modal arbitration.

\begin{table*}[t]
\centering\footnotesize
\caption{Per-configuration evaluation. Entries are conflict gain
(faithful drop), pp relative to Joint, with 95\% paired-bootstrap CIs (1000
resamples). Bold: best gain per row. \GACL{} wins 18/20.}
\label{tab:e-1}
\setlength{\tabcolsep}{4pt}
\begin{adjustbox}{max width=\textwidth}
\begin{tabular}{llccccccc}
\toprule
Model & Task & Joint conf.\ & AAD (drop) & ACD (drop) & \GACL{} (drop) & $\lambda$ & $\tau_A$ \\
\midrule
Qwen2-Audio-Inst. & AQA  &  2.0$\pm$2.0 & $+10.0_{\pm4.0}$\,($1.5_{\pm1.8}$)  & $+16.0_{\pm5.5}$\,($3.0_{\pm2.2}$)  & $\mathbf{+69.5_{\pm6.2}}$\,($6.5_{\pm3.5}$) & 1.25 & 0.5 \\
Qwen2-Audio-Inst. & VSC  & 11.0$\pm$4.2 & $+28.5_{\pm6.2}$\,($3.5_{\pm2.2}$)  & $+21.5_{\pm5.8}$\,($2.5_{\pm2.0}$)  & $\mathbf{+57.5_{\pm7.2}}$\,($6.5_{\pm3.2}$) & 1.0  & 1.5 \\
Qwen2-Audio-Inst. & SER  &  0.0$\pm$0.0 & $+1.0_{\pm1.2}$\,($0.5_{\pm0.8}$)   & $+2.0_{\pm1.8}$\,($1.0_{\pm1.2}$)   & $\mathbf{+4.5_{\pm2.8}}$\,($3.0_{\pm2.2}$)  & 2.0  & 1.5 \\
Qwen2-Audio-Inst. & ALME & 24.5$\pm$5.8 & $+35.0_{\pm6.8}$\,($0.0_{\pm2.0}$)  & $+41.0_{\pm7.2}$\,($3.0_{\pm3.0}$)  & $\mathbf{+66.5_{\pm6.2}}$\,($1.0_{\pm2.2}$) & 2.0  & 0.5 \\
\midrule
Qwen2.5-Omni      & AQA  & 24.5$\pm$6.0 & $+27.0_{\pm6.0}$\,($3.5_{\pm2.5}$)  & $+38.0_{\pm7.0}$\,($6.0_{\pm3.2}$)  & $\mathbf{+41.0_{\pm6.2}}$\,($4.0_{\pm2.8}$) & 1.0  & 0.5 \\
Qwen2.5-Omni      & VSC  &  2.5$\pm$2.2 & $+47.0_{\pm7.0}$\,($1.5_{\pm1.8}$)  & $+56.0_{\pm7.2}$\,($2.5_{\pm2.2}$)  & $\mathbf{+79.0_{\pm6.0}}$\,($4.5_{\pm2.8}$) & 1.25 & 1.5 \\
Qwen2.5-Omni      & SER  &  0.0$\pm$0.0 & $\mathbf{+11.0_{\pm4.2}}$\,($8.0_{\pm3.5}$) & $+6.5_{\pm3.5}$\,($4.5_{\pm2.8}$) & $+10.0_{\pm4.5}$\,($7.0_{\pm3.5}$) & 0.5 & 0.5 \\
Qwen2.5-Omni      & ALME & 30.5$\pm$6.2 & $+52.0_{\pm7.2}$\,($0.0_{\pm2.0}$)  & $+54.0_{\pm7.5}$\,($2.0_{\pm3.0}$)  & $\mathbf{+66.0_{\pm6.5}}$\,($0.0_{\pm2.0}$) & 2.0  & 0.5 \\
\midrule
Voxtral-Small     & AQA  &  3.0$\pm$2.2 & $+6.5_{\pm3.2}$\,($5.5_{\pm3.0}$)   & $+0.0_{\pm1.5}$\,($1.0_{\pm1.2}$)   & $\mathbf{+22.5_{\pm5.5}}$\,($6.5_{\pm3.2}$) & 1.25 & 0.5 \\
Voxtral-Small     & VSC  &  1.5$\pm$1.8 & $+4.0_{\pm2.8}$\,($0.0_{\pm0.0}$)   & $+1.0_{\pm1.2}$\,($0.0_{\pm0.0}$)   & $\mathbf{+22.5_{\pm5.8}}$\,($0.5_{\pm0.8}$) & 0.5  & 1.0 \\
Voxtral-Small     & SER  &  1.0$\pm$1.2 & $+6.0_{\pm3.2}$\,($5.5_{\pm3.0}$)   & $+2.5_{\pm2.2}$\,($0.5_{\pm0.8}$)   & $\mathbf{+6.0_{\pm3.2}}$\,($1.5_{\pm1.8}$)  & 0.5  & 1.5 \\
Voxtral-Small     & ALME & 15.0$\pm$4.8 & $+39.5_{\pm7.5}$\,($5.5_{\pm3.8}$)  & $+34.5_{\pm7.0}$\,($3.0_{\pm2.5}$)  & $\mathbf{+80.0_{\pm6.0}}$\,($1.0_{\pm1.2}$) & 1.0  & 0.5 \\
\midrule
Qwen3-Omni        & AQA  & 43.0$\pm$7.0 & $+26.5_{\pm6.2}$\,($4.5_{\pm3.0}$)  & $+21.0_{\pm5.8}$\,($3.0_{\pm2.2}$)  & $\mathbf{+44.0_{\pm7.0}}$\,($4.5_{\pm3.0}$) & 1.0  & 0.5 \\
Qwen3-Omni        & VSC  & 41.0$\pm$6.5 & $+32.0_{\pm6.5}$\,($8.0_{\pm3.8}$)  & $+18.0_{\pm5.5}$\,($1.5_{\pm1.8}$)  & $\mathbf{+48.5_{\pm7.2}}$\,($9.0_{\pm3.8}$) & 1.25 & 0.5 \\
Qwen3-Omni        & SER  & 16.5$\pm$5.2 & $\mathbf{+17.5_{\pm5.5}}$\,($15.5_{\pm4.8}$) & $+2.5_{\pm2.8}$\,($0.0_{\pm0.0}$) & $+16.0_{\pm5.2}$\,($8.5_{\pm3.8}$) & 0.5 & 1.0 \\
Qwen3-Omni        & ALME & 33.5$\pm$6.5 & $+35.0_{\pm7.0}$\,($1.5_{\pm2.0}$)  & $+38.0_{\pm7.2}$\,($2.5_{\pm2.5}$)  & $\mathbf{+62.5_{\pm7.0}}$\,($-0.5_{\pm0.8}$) & 1.25 & 0.5 \\
\midrule
Kimi-Audio-Inst.  & AQA  &  5.5$\pm$3.2 & $+17.0_{\pm5.2}$\,($3.5_{\pm2.8}$)  & $+17.0_{\pm5.0}$\,($3.5_{\pm2.8}$)  & $\mathbf{+56.0_{\pm6.2}}$\,($7.5_{\pm4.0}$) & 1.0  & 1.0 \\
Kimi-Audio-Inst.  & VSC  &  4.5$\pm$3.0 & $+11.5_{\pm4.2}$\,($3.0_{\pm2.2}$)  & $+18.0_{\pm5.5}$\,($4.0_{\pm2.8}$)  & $\mathbf{+27.0_{\pm6.2}}$\,($8.0_{\pm3.5}$) & 1.25 & 1.0 \\
Kimi-Audio-Inst.  & SER  &  0.0$\pm$0.0 & $+0.5_{\pm0.8}$\,($3.0_{\pm2.2}$)   & $+0.5_{\pm0.8}$\,($0.0_{\pm0.0}$)   & $\mathbf{+8.0_{\pm3.8}}$\,($8.0_{\pm3.8}$)  & 0.5  & 0.5 \\
Kimi-Audio-Inst.  & ALME & 40.0$\pm$6.0 & $+2.0_{\pm6.0}$\,($2.5_{\pm4.8}$)   & $+12.5_{\pm8.8}$\,($11.0_{\pm7.0}$) & $\mathbf{+47.0_{\pm7.0}}$\,($-2.0_{\pm3.5}$) & 2.0 & 0.5 \\
\bottomrule
\end{tabular}
\end{adjustbox}
\end{table*}

\subsection{Reference-Limited SER Decomposition}
\label{app:where-stops}

\GACL{} factorizes gain into (opportunity) $\times$ (gate activation)
$\times$ (donor reliability) $\times$ (conversion). SER is the low-donor
stress case: $56.6 / 94.2 / 35.3 / 48.6$, with final gain $8.9$
(empirical mean, not the algebraic product). Non-SER tasks show $66.3 / 99.3 /
88.3 / 90.9$, with final gain $52.6$. The bottleneck is donor reliability,
not opportunity discovery or gate activation.

\subsection{Ablations under the faithful-drop budget}
\label{app:budgeted-ablation}

\paragraph{Cumulative ablation.}
\tabref{tab:cumulative} shows each component contributes independently:
$\RA$ alone destabilizes the surface (FF $22$\%); adding
$\Nout$ restores it ($90$\%); tuned $\lambda$ then recovers the rescue gap.

\begin{table}[H]
\centering\scriptsize
\setlength{\tabcolsep}{3pt}
\renewcommand{\arraystretch}{1.05}
\begin{tabular}{lccc@{\quad}ccc}
\toprule
Configuration & Rescue & Drop & FF & $\Delta$R & $\Delta$D & $\Delta$FF \\
\midrule
Naive interp.\ ($\alpha{=}0.55$)   & 38.4 & 4.7 & 70.0 & --       & --       & --       \\
\quad+ bound $\alpha{\le}1$        & 38.6 & 4.6 & 70.0 & $+0.2$   & $-0.1$   & $0.0$    \\
\quad+ $\RA$                       & 38.0 & 2.5 & 22.0 & $-0.6$   & $-2.1$   & \cellcolor{red!12}\textbf{$-48.0$} \\
\quad+ $\Nout$                     & 36.2 & 2.1 & 90.0 & $-1.8$   & $-0.4$   & \textbf{$+68.0$} \\
\quad+ tuned $\lambda$ (\GACL{})   & \textbf{45.8} & \textbf{2.1} & \textbf{90.0} & \textbf{$+9.6$} & $0.0$ & $0.0$ \\
\bottomrule
\end{tabular}
\caption{\textbf{Each component contributes independently.} $\Delta$ is change vs.\
the preceding row. FF: free-form VSC exact-surface preservation.}
\label{tab:cumulative}
\end{table}

\paragraph{Strict budgets.}
\tabref{tab:e-5} sweeps budgets: removing $\RA$ appears safe at $5$\,pp but
has measurable cost at $2.5$\,pp. Hard switch attains higher rescue only with
a $45$\,pp maximum drop.

\begin{table}[H]
\centering\scriptsize
\setlength{\tabcolsep}{3pt}
\begin{tabular}{lcccc}
\toprule
Budget & Variant & Conflict gain & Faith.\ drop & Max drop \\
\midrule
2.5 & \GACL{}     & 43.7 [41.4, 45.9] & 1.8 [1.2, 2.6] & 3.0 \\
2.5 & w/o $\RA$   & 41.1 [38.9, 43.1] & 2.3 [1.6, 3.1] & 4.0 \\
2.5 & w/o $\Nout$ & 44.0 [41.8, 46.3] & 1.8 [1.2, 2.5] & 3.0 \\
\midrule
5.0 & \GACL{}     & 45.8 [43.6, 47.8] & 2.1 [1.4, 2.7] & 4.5 \\
5.0 & w/o $\RA$   & 47.4 [45.2, 49.6] & 2.3 [1.6, 3.0] & 4.0 \\
5.0 & w/o $\Nout$ & 46.1 [43.7, 48.4] & 2.1 [1.4, 2.8] & 4.5 \\
5.0 & Hard switch & 66.5 [64.4, 68.6] & 13.3 [12.0, 14.7] & \textbf{45.0} \\
\bottomrule
\end{tabular}
\caption{Strict-budget summary with bootstrap CIs (1000 resamples),
macro-averaged across short-label configurations.}
\label{tab:e-5}
\end{table}

\paragraph{Targeted $\RA$ effect.}
\label{app:ra_targeted}
\tabref{tab:ra-targeted} shows $\RA$ is \emph{targeted}: it barely changes macro
faithful drop but suppresses drop by $1.5$--$1.9\times$ in the
bottom $\RA$ quartile, where audio confidence is weakest.

\begin{table}[H]
\centering\small
\setlength{\tabcolsep}{6pt}
\begin{tabular}{lcccc}
\toprule
Subset & Budget & w/ $\RA$ & w/o $\RA$ & Ratio \\
\midrule
All      & 2.5 & 2.2 & 2.2 & $1.0\times$ \\
Low-$\RA$ & 2.5 & 1.6 & \textbf{3.0} & \textbf{$1.9\times$} \\
All      & 5.0 & 3.1 & 3.5 & $1.1\times$ \\
Low-$\RA$ & 5.0 & 2.1 & \textbf{3.2} & \textbf{$1.5\times$} \\
\bottomrule
\end{tabular}
\caption{$\RA$ acts where it matters. Low-$\RA$ = bottom audio-reliability
quartile within each model--task on faithful samples.}
\label{tab:ra-targeted}
\end{table}

\paragraph{Per-configuration ablation.}
\tabref{tab:e-4} shows hard switch gains raw rescue but creates a large
faithful-drop tail on several configurations.

\begin{table}[H]
\centering\scriptsize
\setlength{\tabcolsep}{3.5pt}
\begin{tabular}{lcccc}
\toprule
Config (gain / drop)         & Hard switch & w/o $\RA$ & w/o $\Nout$ & \GACL{} \\
\midrule
Qwen2-Audio$\times$ALME & 58.0 / 1.0  & 50.5 / 0.5 & 43.0 / 0.5 & 43.0 / 0.5 \\
Qwen2.5$\times$ALME     & 57.5 / 0.5  & 53.0 / 0.5 & 56.0 / 0.5 & 54.0 / 0.5 \\
Voxtral$\times$ALME     & 63.5 / 1.0  & 53.0 / 1.0 & 42.5 / 0.5 & 42.0 / 0.5 \\
Qwen2-Audio$\times$AQA  & 84.0 / 13.5 & 60.5 / 3.5 & 48.5 / 2.5 & 48.5 / 2.5 \\
Qwen2.5$\times$AQA      & 67.0 / 6.5  & 52.5 / 2.5 & 53.5 / 2.0 & 53.0 / 2.0 \\
Voxtral$\times$AQA      & 66.0 / 29.5 & 10.5 / 4.0 & 29.0 / 4.5 & 29.0 / 4.5 \\
Qwen2-Audio$\times$VSC  & 72.0 / 13.5 & 57.5 / 3.0 & 58.0 / 3.0 & 58.0 / 3.0 \\
Qwen2.5$\times$VSC      & 77.0 / 9.5  & 52.0 / 2.5 & 49.5 / 2.5 & 49.5 / 2.5 \\
Voxtral$\times$VSC      & 53.5 / 45.0 & 37.0 / 3.0 & 35.0 / 2.5 & 35.0 / 2.5 \\
\bottomrule
\end{tabular}
\caption{Per-configuration gain / drop. Hard switch's faithful-drop tail
reaches $45$\,pp.}
\label{tab:e-4}
\end{table}

\subsection{Free-form generation}
\label{app:free-form-generation}

\paragraph{Component contribution (VSC).}
The table isolates components on free-form VSC.
Removing $\Nout$ preserves canonical accuracy but rewrites most surface forms
(Exact $22\%$); the disagreement gate protects faithful
surfaces.

\begin{table}[H]
\centering\scriptsize
\setlength{\tabcolsep}{2.5pt}
\begin{tabular}{lcccccc}
\toprule
Variant & Rescue & Drop & Parse fail & Other & Exact & Canon. \\
\midrule
Hard switch  & 90.0 & 10.0 & 0.0 & 10.0 & ---  & ---  \\
w/o $\RA$    & 90.0 & 10.0 & 0.0 & 10.0 & 90.0 & 90.0 \\
w/o $\Nout$  & 88.0 &  8.0 & 0.0 &  9.0 & \cellcolor{red!12}\textbf{22.0} & 92.0 \\
\GACL{}      & 88.0 &  8.0 & 0.0 &  9.0 & \textbf{90.0} & 92.0 \\
\bottomrule
\end{tabular}
\caption{Free-form VSC. Hard switch lacks a logit-interpolation surface, so
exact/canonical are undefined.}
\label{tab:e-6-free-form-vsc}
\end{table}

\paragraph{Cross-task transfer.}
With closed-set hyperparameters, \GACL{}
transfers to free-form AQA, VSC, and ALME with $87$--$100\%$ exact
preservation and near-zero parse failure. SER again acts as the
reference-limited stress case: the audio reference fails to canonicalize
$36/200$ free-form inputs
(\tabref{tab:e-8-parse-failure-counts}), so the donor that \GACL{}
interpolates toward is unreliable on $18\%$ of samples, the free-form
analogue of the closed-set regime in \secref{sec:component}.

\begin{table}[H]
\centering\scriptsize
\setlength{\tabcolsep}{3pt}
\begin{tabular}{lccccc}
\toprule
Task & Rescue & Drop & Parse fail & Exact & Canon. \\
\midrule
AQA  & 75.5 & 12.0 & 0.0 & 87.4 & 87.4 \\
VSC  & 88.0 &  8.0 & 0.0 & 90.0 & 92.0 \\
ALME & 67.0 & $-0.5$ & 1.2 & 100.0 & 100.0 \\
SER$^\dagger$ & 19.5 & 55.0 & 0.0 & 43.0 & 45.0 \\
\bottomrule
\end{tabular}
\caption{Cross-task free-form transfer on Qwen2-Audio-Inst.\ with
closed-set hyperparameters. $^\dagger$SER's audio reference
canonicalizes only $82\%$ of inputs (\tabref{tab:e-8-parse-failure-counts}),
making the setting reference-limited.}
\label{tab:e-7-cross-task-free-form}
\end{table}

\begin{table}[H]
\centering\scriptsize
\caption{Free-form parse failures (out of 200) on
Qwen2-Audio-Inst.}
\label{tab:e-8-parse-failure-counts}
\setlength{\tabcolsep}{6pt}
\begin{tabular}{lcccc}
\toprule
Branch          & AQA & VSC & SER & ALME \\
\midrule
Audio reference & 0 & 1 & \textbf{36} & 4 \\
\GACL{}         & 0 & 0 & 0  & 5 \\
\bottomrule
\end{tabular}
\end{table}

\paragraph{Bound ablation under SER stress.}
\tabref{tab:e-9-pure-bound-ablation} isolates the upper cap $\alphamax=1$:
removing it produces fluent but invalid generations when extrapolation
crosses the audio-reference endpoint. The Would exceed column counts
samples \GACL{} clips at $\alpha=1$. Without the cap,
clipped samples produce $26$--$30\%$ parse failures, confirming the cap
protects output validity in the reference-limited regime.

\begin{table}[H]
\centering\scriptsize
\setlength{\tabcolsep}{2.2pt}
\begin{adjustbox}{max width=\columnwidth}
\begin{tabular}{llcccccc}
\toprule
\multirow{2}{*}{Regime} & \multirow{2}{*}{Variant} & \multirow{2}{*}{Acc} & \multirow{2}{*}{Parse fail} & \multirow{2}{*}{Rescue / drop} & Realized & \multicolumn{2}{c}{Counterfactual w/o bound} \\
\cmidrule(lr){7-8}
 & & & & & $\alpha>1$ & Would exceed & Parse fail \\
\midrule
\multirow{2}{*}{Faithful} & bounded   & 45.0 & 0.0  & 55.0 & 0.0  & 54.0 & 0.0  \\
                          & w/o bound & 45.0 & 16.0 & 55.0 & 54.0 & 54.0 & 29.6 \\
\midrule
\multirow{2}{*}{Conflict} & bounded   & 19.5 & 0.0  & 19.5 & 0.0  & 61.5 & 0.0  \\
                          & w/o bound & 14.0 & 16.5 & 14.0 & 61.5 & 61.5 & 26.8 \\
\bottomrule
\end{tabular}
\end{adjustbox}
\caption{SER free-form bound ablation. The cap enforces output validity
rather than canonical-accuracy gain.}
\label{tab:e-9-pure-bound-ablation}
\end{table}

\begin{table}[H]
\centering\scriptsize
\setlength{\tabcolsep}{2.2pt}
\begin{adjustbox}{max width=\columnwidth}
\begin{tabular}{lcccll}
\toprule
Regime & ID & Gold & Text & Bounded & w/o bound \\
\midrule
Faith. & ser\_0006 & happy   & happy   & \emph{Angry mood} (angry)              & \emph{Every day.} ($\emptyset$) \\
Faith. & ser\_0527 & neutral & neutral & \emph{Surprised and disgusted} (disgusted) & \emph{sound of head impact} ($\emptyset$) \\
Conf.  & ser\_0305 & neutral & happy   & \emph{The emotion is neutral.} (neutral)   & \emph{the critical moment} ($\emptyset$) \\
Conf.  & ser\_0527 & neutral & fearful & \emph{Surprised and disgusted} (disgusted) & \emph{Off the top of his head} ($\emptyset$) \\
\bottomrule
\end{tabular}
\end{adjustbox}
\caption{Representative SER extrapolation risks. Without the bound,
generation drifts out of the SER label space rather than to a near-miss
emotion, confirming the cap protects validity, not accuracy.}
\label{tab:e-10-ser-extrapolation-failures}
\end{table}

\subsection{Hard selection baseline}
\label{app:hard-selection}

Budgeted reference selection (BRS) sets $\hat y(x)=\hat y_A$ if
$\Delta_A(x)>\theta$, else $\hat y_J$, and sweeps $\theta$ for a frontier
comparable to nAUC. Macro results: at @0--5, BRS $29.5$ vs.\ \GACL{} $33.3$ ($+3.8$;
\GACL{} wins 14/20); at @0--10, $36.6$ vs.\ $39.2$ ($+2.6$; 14/20). BRS
approaches \GACL{} when audio-reference reliability is uniformly high
(e.g., ALME) and lags when the reference is informative but imperfect.

\begin{table}[H]
\centering\scriptsize
\setlength{\tabcolsep}{2.5pt}
\begin{adjustbox}{max width=\columnwidth}
\begin{tabular}{llccc@{\quad}ccc}
\toprule
\multirow{2}{*}{Model} & \multirow{2}{*}{Task}
 & \multicolumn{3}{c}{nAUC@0--5} & \multicolumn{3}{c}{nAUC@0--10} \\
 &  & BRS & \GACL{} & $\Delta$ & BRS & \GACL{} & $\Delta$ \\
\midrule
\multirow{4}{*}{Qwen2-Audio-Inst.} & AQA  & 41.3 & \textbf{51.0} & $+9.6$  & 55.3 & \textbf{60.7} & $+5.4$ \\
                                   & VSC  & 16.7 & \textbf{39.1} & $+22.4$ & 36.0 & \textbf{51.2} & $+15.2$ \\
                                   & SER  &  2.7 & \textbf{2.9}  & $+0.2$  &  5.3 & \textbf{6.1}  & $+0.8$ \\
                                   & ALME & \textbf{64.2} & 60.4 & $-3.9$ & \textbf{66.1} & 63.4 & $-2.7$ \\
\cmidrule(lr){2-8}
\multirow{4}{*}{Qwen2.5-Omni}      & AQA  & 22.6 & \textbf{31.6} & $+8.9$ & 42.4 & \textbf{45.7} & $+3.3$ \\
                                   & VSC  & 57.8 & \textbf{58.0} & $+0.3$ & 70.4 & \textbf{71.6} & $+1.2$ \\
                                   & SER  &  7.5 & \textbf{7.6}  & $+0.1$ & 11.0 & \textbf{11.4} & $+0.4$ \\
                                   & ALME & 65.0 & \textbf{66.0} & $+1.0$ & 65.0 & \textbf{66.0} & $+1.0$ \\
\cmidrule(lr){2-8}
\multirow{4}{*}{Voxtral-Small}     & AQA  & \textbf{11.6} & 8.8 & $-2.8$ & \textbf{19.1} & 17.4 & $-1.6$ \\
                                   & VSC  & 14.7 & \textbf{25.0} & $+10.4$ & 19.6 & \textbf{28.1} & $+8.5$ \\
                                   & SER  &  2.7 & \textbf{6.7}  & $+4.0$  &  6.6 & \textbf{10.4} & $+3.8$ \\
                                   & ALME & \textbf{72.8} & 72.0 & $-0.8$ & \textbf{76.9} & 76.0 & $-0.9$ \\
\cmidrule(lr){2-8}
\multirow{4}{*}{Qwen3-Omni}        & AQA  & \textbf{30.7} & 30.1 & $-0.7$ & \textbf{38.8} & 38.6 & $-0.1$ \\
                                   & VSC  & 21.4 & \textbf{36.2} & $+14.8$ & 34.7 & \textbf{42.5} & $+7.7$ \\
                                   & SER  &  4.5 & \textbf{6.6}  & $+2.0$  &  7.2 & \textbf{10.2} & $+3.0$ \\
                                   & ALME & 63.5 & \textbf{64.0} & $+0.5$ & 63.5 & \textbf{64.0} & $+0.5$ \\
\cmidrule(lr){2-8}
\multirow{4}{*}{Kimi-Audio-Inst.}  & AQA  & 26.0 & \textbf{39.3} & $+13.3$ & 38.4 & \textbf{48.7} & $+10.3$ \\
                                   & VSC  & \textbf{14.6} & 12.7 & $-1.9$ & \textbf{23.3} & 19.2 & $-4.1$ \\
                                   & SER  &  0.6 & \textbf{2.0}  & $+1.3$ &  2.9 & \textbf{4.9}  & $+2.0$ \\
                                   & ALME & \textbf{49.0} & 47.0 & $-2.0$ & \textbf{49.0} & 47.0 & $-2.0$ \\
\bottomrule
\end{tabular}
\end{adjustbox}
\caption{\GACL{} vs.\ budgeted reference selection. \GACL{} wins 14/20 at
both budgets; losses concentrate on ALME (uniformly high $\RA$, where
BRS suffices) and a few high-baseline configurations.}
\label{tab:e-11}
\end{table}

\subsection{Prompt-level interventions}
\label{app:prompt-interventions}

We evaluate four prompt variants on Qwen2-Audio-Inst.\ and Qwen2.5-Omni
across AQA, VSC, and SER (\tabref{tab:prompt_baselines}). The strongest
variant, audio prioritization, gains $+6.7$\,pp conflict audio-follow over
the default; \GACL{} at matched faithful drop gains $+34.1$\,pp. Prompt
instructions bias generation but do not intervene on arbitration
computation.

\begin{table}[H]
\centering\small
\setlength{\tabcolsep}{4pt}
\begin{adjustbox}{max width=\columnwidth}
\begin{tabular}{lccc}
\toprule
Variant & Audio-follow $\uparrow$ & $\Delta$ & Drop \\
\midrule
Original prompt   & 15.0 & --   & 0.0 \\
Audio-priority    & 21.7 & $+6.7$ & 0.9 \\
Ignore-text       & 16.7 & $+1.7$ & 1.0 \\
Bias-awareness    & 18.6 & $+3.6$ & 0.9 \\
Chain-of-thought  & 17.2 & $+2.2$ & 1.1 \\
\midrule
\textbf{\GACL{} (matched drop)} & \textbf{49.1} & \textbf{$+34.1$} & \textbf{0.9} \\
\bottomrule
\end{tabular}
\end{adjustbox}
\caption{Prompt-level interventions vs.\ \GACL{} at matched faithful drop.
The \GACL{} gain is $5\times$ the best fixed-prompt gain.}
\label{tab:prompt_baselines}
\end{table}

\subsection{Fine-tuning baseline setup}
\label{app:sft-details}

The LoRA baseline in the main text uses Qwen2-Audio-7B-Instruct
with rank $r{=}8$, $\alpha{=}32$, applied to seven LLM projections (Q, K,
V, O, gate, up, down); the audio encoder is frozen. Training data include $1{,}000$
MCR-Bench base samples (AQA $500$ + SER $500$), paired into faithful
and adversarial forms ($2{,}000$ total) at $1{:}1$. Optimization uses $4\times$A800
with DDP, lr $10^{-4}$, $2$ epochs, effective batch $128$, and AdamW. The
MCR-Bench split convention is preserved, with no test-set leakage. Evaluation
uses the same $200$-example conflict test split throughout.

\end{document}